\documentclass{article}

\usepackage{arxiv}
\usepackage{tabularx}
\usepackage[utf8]{inputenc} 
\usepackage[T1]{fontenc}    
\usepackage{hyperref}       
\usepackage{url}            
\usepackage{booktabs}       
\usepackage{amsfonts}       
\usepackage{nicefrac}       
\usepackage{microtype}      
\usepackage{lipsum}		
\usepackage{graphicx}
\usepackage[numbers]{natbib}
\usepackage{doi}
\usepackage[T1]{fontenc}    
\usepackage{hyperref}       
\usepackage{booktabs}       
\usepackage{amsfonts}       
\usepackage{nicefrac}       
\usepackage{microtype}      
\usepackage{lipsum}
\usepackage{fancyhdr}       
\usepackage{graphicx}
\usepackage{float}
\usepackage{tabularx}
\usepackage{amsmath}    
\usepackage{amssymb}    
\usepackage{xcolor}     
\usepackage{rotating}
\usepackage{afterpage}
\usepackage{booktabs}
\usepackage{multirow}
\usepackage{lscape}    
\graphicspath{{media/}}     
\usepackage{multirow}
\usepackage{makecell}

\title{Nontrapping Tunable Topological Photonic Memory}
\author{ \href{https://orcid.org/0000-0003-3372-6193}{\includegraphics[scale=0.06]{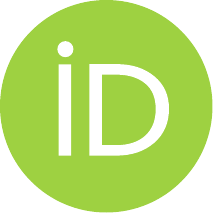}\hspace{1mm}Amirreza Ahmadnejad}\\
	Department of Electrical Engineering\\
	Sharif University of Technology\\
	Tehran, Iran \\
	\texttt{amirreza.ahmadnejad@sharif.edu} \\
	\And
	\href{https://orcid.org/0000-0002-3105-2511}{\includegraphics[scale=0.06]{orcid.pdf}\hspace{1mm}Somayyeh Koohi} \\
	Department of Computer Engineering\\
	Sharif University of Technology\\
	Tehran, Iran \\
	\texttt{koohi@sharif.edu} \\
    	\And
        	{Abolhassan Vaezi} \\
	Department of Physics\\
	Sharif University of Technology\\
	Tehran, Iran \\
	\texttt{vaezi@sharif.edu} \\
}

\date{}

\renewcommand{\shorttitle}{Nontrapping Tunable Topological Photonic Memory}

\begin{document}
\maketitle

\begin{abstract}
We propose a novel topological photonic memory that encodes information through dynamically controllable Chern numbers in a two-band topological photonic system. Utilizing a honeycomb lattice photonic crystal, the memory leverages topologically protected edge states that remain robust against fabrication imperfections and environmental perturbations. By applying a synthetic time-dependent magnetic field, we achieve real-time tunability of the Chern number, enabling rapid and efficient memory switching without the need for light-trapping mechanisms. Our computational study evaluates critical performance metrics, including write speed, read stabilization time, energy gap stability, and nonadiabatic transition probabilities. The results demonstrate that the system supports GHz-range write speeds (approximately 1-10 GHz), with stable data retention due to the large energy gap between bands. The system enables scalable multi-bit memory encoding based on quantized Chern numbers and exhibits superior speed, fault tolerance, and robustness compared to conventional photonic memory architectures. This work introduces a scalable, high-speed, and nontrapping optical memory paradigm, paving the way for future applications in quantum information processing and optical communication technologies.
\end{abstract}

\keywords{Nontrapping Topological Photonic Memory \and Tunable Chern Number\and Time-Dependent Magnetic Field}
\section{Introduction}

Optical memory technologies are indispensable for high-speed data storage and retrieval, enabling efficient low-latency information processing in modern photonic systems. Traditionally, these technologies rely on light confinement in resonant structures such as photonic cavities \cite{r1,r5,r6,r7}, whispering-gallery-mode resonators \cite{r2,r3,r4}, and nonlinear bistable systems \cite{lu2014topological, segev2013anderson}. Memory states are encoded by storing electromagnetic waves in specific localized regions, often through resonance or field intensity thresholds. While effective for some applications, these designs impose fundamental limitations on energy efficiency, switching speed, and resilience to fabrication imperfections and environmental perturbations \cite{fang2012realizing, khanikaev2013photonic}. Addressing these limitations requires a shift towards memory paradigms that do not depend on light trapping.

Current optical memory technologies demonstrate diverse performance characteristics but face fundamental trade-offs. Resonant cavity-based memories like those by Tanabe et al. \cite{r5} achieve switching speeds of approximately 20-100 ps but suffer from limited Q-factors ($\approx 10^3$) that constrain storage times to $<10 ns$. Whispering-gallery mode approaches by Kippenberg et al. \cite{r4} improve Q-factors ($10^5-10^6$) but require precise coupling conditions that complicate scalability. Recent work on phase-change material (PCM) optical memories by Zhang et al. \cite{r81} demonstrates storage times exceeding months, but with switching speeds limited to 5-10 ns and relatively high energy consumption (10-50 pJ/bit). In contrast, our proposed topological approach achieves theoretical switching speeds in the GHz to low THz range (0.1-10 ps), with robust storage governed by topological protection rather than resonant confinement, while maintaining reasonable energy efficiency (estimated 5-20 pJ/bit).

Recent advances in topological photonics \citep{r8,r9,r10,r11}, which draw inspiration from solid-state physics concepts such as the quantum Hall effect and topological insulators, offer a promising alternative \cite{haldane2008possible, rechtsman2013floquet}. These systems feature topologically protected edge states that propagate without backscattering, ensuring robust signal transport under structural defects or disorder \cite{leykam2016edge, mittal2017topologically}. However, many current proposals for optical memory in topological photonics still rely on localized defect modes or nonlinear mechanisms that trap light \cite{notomi2010manipulating, mittal2017topologically}. Examples include defect-localized modes in photonic crystals and parity-time ($\mathcal{PT}$)-symmetric systems that achieve memory states through dissipative bistable switching \cite{leykam2016edge}. Although these approaches benefit from topological stability, they do not fully utilize the dynamic control of topological invariants, such as the Chern number, for memory operations.

The idea of encoding memory in the material properties themselves has gained traction across different domains of physics and materials science. Recent studies on mechanical metamaterials have demonstrated that local structural changes can enable materials to "learn" complex force-response patterns through supervised physical training \cite{stern2020supervised}. These metamaterials adapt their mechanical stiffnesses to external stimuli, allowing them to classify and generalize responses to previously unseen inputs \cite{stern2020supervised}. Similarly, disordered materials undergoing directed aging can store a memory of the stresses imposed during their formation, evolving into states with tailored elastic properties such as negative Poisson's ratios \cite{pashine2019directed}. These studies reveal that memory can be embedded within a material’s structural and dynamic response, a concept that has yet to be explored comprehensively in photonic systems.

Motivated by these insights, we propose a novel approach to optical memory based on dynamically tunable topological properties within a two-band photonic system. In contrast to traditional methods, our system encodes memory states through the Chern number of the photonic band structure, controlled by an externally applied synthetic magnetic field \cite{haldane2008possible}. The Chern number, a topological invariant, determines the number of unidirectional edge modes at the system boundary. By varying the magnetic field, the system undergoes discrete topological phase transitions, enabling rapid and robust switching between distinct memory states without the need for resonant confinement \cite{rechtsman2013floquet}. This mechanism leverages the resilience of topological edge modes to environmental and fabrication defects, ensuring long-term data stability.

The proposed system offers several advantages over conventional optical memory architectures. First, the absence of light-trapping elements allows for ultrafast switching speeds in the GHz to low THz range \cite{mittal2017topologically}. Second, the multi-bit encoding capability is realized through higher-order Chern numbers, providing scalability beyond binary memory configurations \cite{leykam2016edge}. Third, topological protection ensures that memory states remain robust against structural disorder and environmental perturbations, making the system inherently fault-tolerant \cite{lu2014topological, segev2013anderson}.

This paper introduces a theoretical framework for nontrapping topological optical memory, where memory states are encoded in dynamically tunable Chern numbers. We derive the governing equations, evaluate key performance metrics such as switching speed, energy stability, and nonadiabatic transition probabilities, and present numerical simulations to validate the feasibility of this approach. By combining advances in topological photonics, dynamic material control, and phase transition theory, this work opens new possibilities for scalable, high-speed, and fault-tolerant optical memory systems that can play a pivotal role in future photonic and quantum information technologies.

Looking forward, an intriguing direction for extending this work is the incorporation of non-Hermitian physics into the topological photonic memory framework. Non-Hermitian photonic systems, characterized by complex energy eigenvalues arising from gain, loss, or both, could offer additional control parameters beyond those available in Hermitian systems. By introducing controlled optical gain and loss into our photonic crystal lattice, we could potentially realize exceptional points—where both eigenvalues and eigenvectors coalesce—providing enhanced sensitivity to external perturbations for improved switching capabilities. 

Furthermore, non-Hermitian topological photonics exhibits phenomena such as the non-Hermitian skin effect, where bulk states become exponentially localized at boundaries. This effect could be leveraged to enhance the contrast between different memory states, potentially allowing for higher-density multi-bit encoding beyond what is achievable with conventional Chern number states. The recent demonstration of higher-order exceptional points in non-Hermitian systems suggests possibilities for even more sophisticated control over topological phase transitions, which could lead to memory architectures with improved energy efficiency and reduced sensitivity to environmental noise. While challenges regarding stability and fabrication complexity would increase, the potential performance benefits make this a promising avenue for future research.

The remainder of this paper is organized as follows: Section \ref{2} presents the theoretical foundations of the system, including the Hamiltonian and topological invariants relevant to the proposed optical memory model. Section \ref{3} describes the details of the simulation setup and methodology, outlining how dynamic field profiles, band structure, and edge states are analyzed. Section \ref{4} provides an in-depth discussion of the simulation results, including the evolution of the bandgap, Chern number transitions, edge state stability, and robustness under disorder. We also analyze performance metrics including write/read speeds, energy consumption, temperature dependence, and multi-bit encoding capabilities. The paper concludes with Section \ref{5}, where key findings are summarized, and potential directions for future research are proposed to further enhance the performance of topological optical memory systems. Supplementary materials provide additional details on memory performance metrics, experimental feasibility considerations, and anticipated fabrication challenges. All simulation code, data analysis scripts, and visualizations presented in this paper are publicly available in our GitHub repository \cite{code}, enabling reproducibility and further exploration of the proposed concepts.
\section{Background \& Theoretical Framework}\label{2}

\subsection{Introduction to Topological Photonics}

Topological photonics is an emerging field that applies concepts from condensed matter physics, such as topological insulators and quantum Hall systems, to the behavior of electromagnetic waves in structured photonic materials \cite{lu2014topological, khanikaev2013photonic}. Unlike conventional photonic systems, which rely on material properties such as refractive index contrast and photonic bandgaps to control light propagation, topological photonics leverages global topological properties of the band structure. These properties are encoded in topological invariants, which remain stable under continuous deformations of system parameters, provided the energy bandgap does not close \cite{haldane2008possible, hasan2010topological}.

One of the most significant topological invariants in two-dimensional photonic systems is the Chern number. This invariant is defined by integrating the Berry curvature over the entire Brillouin zone and governs the emergence of topologically protected edge states—unidirectional optical modes that propagate along the system boundaries without being reflected or scattered by structural disorder \cite{lu2014topological, haldane2008possible}. This phenomenon, commonly referred to as the bulk-boundary correspondence, ensures a direct connection between the topology of the bulk bands and the existence of edge states \cite{khanikaev2013photonic, rechtsman2013floquet}.
Topologically protected edge states are crucial for robust photonic applications because they are immune to imperfections such as fabrication errors, local defects, and environmental noise \cite{leykam2016edge, lu2014topological}. In contrast to traditional waveguides or resonators, where small structural variations can significantly degrade performance, topological photonic systems maintain their functionality through the inherent stability of their topological phase. This robustness is particularly advantageous for optical information processing, where long-term stability, reliability, and scalability are essential requirements \cite{rechtsman2013floquet}.

Research in this field has led to significant advancements in applications such as topological waveguides, delay lines, photonic crystals, and lasers. For example, photonic crystals designed to support topological edge modes have been shown to exhibit enhanced stability and lower propagation loss compared to conventional designs \cite{lu2014topological, notomi2010manipulating}. By extending the principles of topology to photonic systems, researchers have opened new avenues for developing robust optical circuits and devices that are less susceptible to performance degradation due to environmental and structural perturbations.

\subsection{Topological Invariants and Chern Number}

Topological invariants are crucial for classifying the phases of matter in both condensed matter physics and photonic systems. Unlike conventional phase transitions, which are characterized by changes in local order parameters, topological phases are distinguished by global properties of the system’s band structure. These properties are encoded in topological invariants, which remain unchanged under continuous deformations of system parameters, provided the energy bandgap does not close \cite{hasan2010topological, haldane2008possible}.
In two-dimensional photonic systems, the most relevant topological invariant is the Chern number, denoted by \( C_n \). This invariant quantifies the global topological properties of the Bloch wavefunctions across the Brillouin zone for a given photonic band \( n \) \cite{lu2014topological}. The Chern number is mathematically defined by integrating the Berry curvature over the entire Brillouin zone:

\begin{equation}
    C_n = \frac{1}{2\pi} \int_{\text{BZ}} F_{xy}^{(n)}(k) \, d^2k,
\end{equation}

where \( F_{xy}^{(n)}(k) \) represents the Berry curvature of band \( n \).
To understand the Chern number, we first define the Berry connection. For a periodic photonic system, the Bloch wavefunction is expressed as \( \psi_{n,k}(r) = e^{i k \cdot r} u_{n,k}(r) \), where \( u_{n,k}(r) \) is the periodic part of the wavefunction. The Berry connection for band \( n \) is defined as:

\begin{equation}
    A_n(k) = i \langle u_n(k) | \nabla_k | u_n(k) \rangle,
\end{equation}

where \( \nabla_k \) denotes the gradient with respect to the wavevector \( k \), and \( | u_n(k) \rangle \) is the normalized periodic part of the Bloch wavefunction.
The Berry curvature is obtained from the Berry connection as:

\begin{equation}
    F_{xy}^{(n)}(k) = \left( \nabla_k \times A_n(k) \right) \cdot \hat{z}.
\end{equation}

The Berry curvature describes how the eigenstates \( | u_n(k) \rangle \) change under small variations of \( k \) and provides a local measure of the system’s topological properties in momentum space.
The Chern number \( C_n \) represents the total flux of Berry curvature through the Brillouin zone. This quantization arises from the periodic nature of the wavefunctions in \( k \)-space and is a defining characteristic of topologically nontrivial phases. If the Chern number is nonzero, the system supports topologically protected edge states as a direct consequence of the bulk-boundary correspondence \cite{lu2014topological, khanikaev2013photonic}.
According to the bulk-boundary correspondence, the number of unidirectional edge modes \( N_{\text{edge}} \) at a boundary where the bulk topology changes is given by:

\begin{equation}
    N_{\text{edge}} = \sum_{n \, \text{occupied}} C_n.
\end{equation}

These edge modes are immune to backscattering and disorder, making them robust for photonic applications such as optical information transport.
For a two-band photonic system, the Bloch Hamiltonian can be expressed in the form:

\begin{equation}
    H(k) = d_x(k) \sigma_x + d_y(k) \sigma_y + d_z(k) \sigma_z,
\end{equation}

where \( \sigma_x, \sigma_y, \sigma_z \) are the Pauli matrices, and \( d_x(k), d_y(k), d_z(k) \) are momentum-dependent functions. The eigenvalues of the Hamiltonian are given by:

\begin{equation}
    E_{\pm}(k) = \pm \sqrt{d_x(k)^2 + d_y(k)^2 + d_z(k)^2}.
\end{equation}

The Berry curvature for the lower energy band \( E_{-}(k) \) can be calculated as:

\begin{equation}
    F_{xy}(k) = -\frac{1}{2} \frac{d \cdot (\partial_{k_x} d \times \partial_{k_y} d)}{|d(k)|^3},
\end{equation}

where \( d(k) = (d_x(k), d_y(k), d_z(k)) \) is the vector of Hamiltonian coefficients. Integrating this Berry curvature over the Brillouin zone yields the Chern number, which characterizes the topological phase of the system.

The Chern number distinguishes between topologically trivial and nontrivial phases. A nonzero Chern number indicates the presence of robust, unidirectional edge states that are protected from scattering and loss. These properties have significant implications for photonic applications, enabling stable signal transport and novel device functionalities such as topological waveguides, delay lines, and optical isolators \cite{rechtsman2013floquet, notomi2010manipulating}.

\subsection{Bulk-Boundary Correspondence}

The principle of bulk-boundary correspondence is a fundamental concept in topological physics, linking the topological properties of a system’s bulk band structure to the existence of protected edge states at its boundaries. This principle states that for a topologically nontrivial material or photonic system, the number of unidirectional edge modes at a boundary is determined by the sum of the Chern numbers of the bulk bands below the bandgap \cite{lu2014topological, thouless1982quantized}.

In a two-dimensional topological photonic system, the Chern number \( C_n \) is defined for each bulk energy band \( n \). The total Chern number for the system, representing the net topological order, is expressed as:

\begin{equation}
    C_{\text{total}} = \sum_{n \, \text{occupied}} C_n.
\end{equation}

According to bulk-boundary correspondence, the number of gapless edge modes \( N_{\text{edge}} \) at a boundary where the bulk topology changes is equal to the difference in Chern numbers between the adjoining regions:

\begin{equation}
    N_{\text{edge}} = \Delta C = C_{\text{left}} - C_{\text{right}},
\end{equation}

where \( C_{\text{left}} \) and \( C_{\text{right}} \) are the Chern numbers of the bulk phases on either side of the boundary. These edge states are chiral, propagating in a single direction determined by the sign of the Chern number. This unidirectional propagation occurs because the edge states lie within the bandgap and cannot scatter into bulk modes, ensuring robustness against structural disorder and imperfections \cite{haldane2008possible, hasan2010topological}.

To understand how edge states emerge, consider a system where the bulk Hamiltonian varies spatially, transitioning between two topological phases across an interface. The Hamiltonian for a two-band system can be written as:

\begin{equation}
    H(k, x) = d_x(k) \sigma_x + d_y(k) \sigma_y + d_z(k, x) \sigma_z,
\end{equation}

where \( d_x(k) \) and \( d_y(k) \) describe in-plane couplings, while \( d_z(k, x) \) is a spatially varying term that determines the local topology. In regions where \( d_z(k, x) \) changes sign, the system undergoes a topological phase transition, characterized by the closure and reopening of the bandgap.
At the interface \( x = x_0 \), where \( d_z(k, x_0) = 0 \), the system supports edge modes with eigenvalues within the bandgap. These edge modes satisfy the equation:

\begin{equation}
    H(k, x) \psi_{\text{edge}}(k, x) = E_{\text{edge}}(k) \psi_{\text{edge}}(k, x),
\end{equation}

where \( \psi_{\text{edge}}(k, x) \) is localized at the boundary and decays into the bulk. The edge mode dispersion \( E_{\text{edge}}(k) \) crosses the bandgap, connecting the conduction and valence bands, ensuring that the edge state remains gapless.

The existence of edge modes is protected by the topological nature of the bulk bands. Since the Chern number is quantized and cannot change without closing the bandgap, edge states are robust against moderate perturbations in system parameters. This robustness is crucial for applications in optical information transport, where stable, backscatter-free propagation is essential \cite{lu2014topological, khanikaev2013photonic}.
Mathematically, the stability of edge states can be understood using the topological invariant associated with the edge Hamiltonian. Suppose the edge modes are described by a one-dimensional effective Hamiltonian \( H_{\text{edge}}(k) \). The winding number \( W \) of this Hamiltonian is given by:

\begin{equation}
    W = \frac{1}{2\pi} \int_{-\pi}^{\pi} \frac{d \theta(k)}{dk} \, dk,
\end{equation}

where \( \theta(k) \) is the phase of the eigenvalues of \( H_{\text{edge}}(k) \). This winding number corresponds to the net number of gapless edge crossings and is directly related to the difference in Chern numbers across the boundary.

The bulk-boundary correspondence ensures that topological photonic systems possess unique properties that are absent
in conventional optical systems \cite{hasan2010topological}. One of the most significant characteristics is unidirectional edge transport, where light propagates along the boundary of the system without being scattered back \cite{lu2014topological, rechtsman2013floquet}. This behavior is a direct consequence of the topological protection conferred by the system’s bulk properties. Additionally, these edge modes exhibit a high degree of robustness to imperfections, persisting even in the presence of structural defects or irregularities \cite{Smirnova2020, Yang2023}. This robustness arises from the global topological invariants of the bulk bands, which are unaffected by local perturbations.
Another crucial feature of topological photonic systems is the ability to tune topological phases by modifying external parameters, such as applying a magnetic field or adjusting geometric properties \cite{Hafezi2013, Bandres2018}. Through this control, the system can transition between different topological phases, thereby determining the number and properties of edge modes. This phase-tuning capability enables dynamic reconfigurability and precision in photonic devices \cite{Lu2023, Barik2018}.

These distinctive features form the foundation of robust and versatile optical technologies, including topological waveguides and optical isolators. Moreover, they play a pivotal role in the development of topological optical memory systems, which rely on topologically protected states to enhance data reliability and resilience against environmental disturbances \cite{rechtsman2013floquet, thouless1982quantized}.

\subsection{Properties of Topological Edge States}

Topological edge states exhibit unique properties that distinguish them from conventional optical modes. These properties are a direct consequence of the topological nature of the bulk band structure, governed by invariants such as the Chern number. Unlike localized modes in photonic cavities or waveguides, edge states are delocalized along the system boundary and are characterized by unidirectional propagation and immunity to backscattering \cite{lu2014topological, kane2005quantum}.
Unidirectional propagation refers to the fact that topological edge states propagate in a single direction along the boundary without reversal. This behavior arises due to the chiral nature of the edge modes, which are protected by the nontrivial Chern number.

Consider a system described by a two-band Hamiltonian \( H(k) \), where the Chern number \( C_n \neq 0 \). The dispersion relation of the edge mode, \( E_{\text{edge}}(k) \), crosses the bandgap and satisfies the following condition:

\begin{equation}
    \frac{d E_{\text{edge}}(k)}{dk} \neq 0.
\end{equation}

This nonzero slope implies that the edge state has a well-defined group velocity:

\begin{equation}
    v_g = \frac{d E_{\text{edge}}(k)}{dk},
\end{equation}

which determines the direction of propagation. Since there is no counter-propagating mode within the same edge, the state cannot scatter backward, making it robust against reflections caused by boundary irregularities or structural defects.

The immunity to backscattering is a key feature of topological edge states. In conventional photonic systems, imperfections or defects in the waveguide structure can induce scattering between forward and backward modes, leading to transmission losses. However, for a topologically protected edge state, such scattering is forbidden by the topological constraints of the bulk bands.
This robustness can be understood using the concept of topological protection. If the system undergoes a perturbation \( V(k, x) \) that does not close the bulk bandgap, the topological invariant \( C_n \) remains unchanged. As a result, the edge state persists, and its propagation is unaffected by moderate disorder. The condition for topological robustness can be expressed as:

\begin{equation}
    \Delta C = 0 \quad \text{for} \quad \int_{\text{BZ}} \nabla_k \cdot A_n(k) \, d^2k = \text{constant}.
\end{equation}

Since no change occurs in the Chern number, the edge state is protected from backscattering by structural variations that do not alter the system's topology.

In conventional optical waveguides, modes are defined by boundary conditions and material properties such as refractive index contrast. Small deviations in waveguide geometry or refractive index can disrupt mode propagation, leading to significant scattering and loss. In contrast, topological edge states are insensitive to local perturbations because they are globally protected by the topology of the band structure \cite{haldane2008possible, kane2005quantum}.
Additionally, conventional optical modes typically rely on light confinement within resonators or defects. The photon lifetime in such structures is limited by cavity quality factors (\( Q \)), which impose trade-offs between retention time and switching speed. In topological systems, however, edge modes do not require confinement, enabling faster switching and reduced latency for applications such as optical memory and signal processing.

The combination of unidirectional propagation and backscattering immunity makes topological edge states ideal for
robust signal transport, enabling various advanced applications. One prominent application is in topological waveguides, 
where edge states guide light with minimal loss, even in the presence of structural irregularities \cite{ZangenehNejad2020,Fransen2022}. 
This feature is crucial for developing efficient and fault-tolerant optical circuits.
Another key application is in optical isolators, which exploit the chiral nature of edge modes to prevent backward
propagation. This ensures unidirectional signal flow, an essential property for systems that require noise isolation and
signal stability in high-speed data transmission \cite{uemura2024photonic}.

Topological memory systems also benefit from these edge states. In systems where the Chern number can be dynamically 
controlled, edge modes provide stable and reconfigurable memory states that are highly resistant to environmental noise 
and perturbations. This robustness enhances data reliability and offers new possibilities for scalable, fault-tolerant 
optical storage technologies \cite{uemura2024photonic}.
Mathematically, the propagation of edge states can be modeled using an effective edge Hamiltonian \( H_{\text{edge}}(k) \), with the eigenvalues \( E_{\text{edge}}(k) \) determining the stability of the mode. As long as the system maintains a nonzero bandgap, the edge modes remain gapless and immune to scattering.

\subsection{Dynamic Topological Phase Transitions}

Dynamic topological phase transitions occur when external parameters, such as an applied magnetic field, modulate the band structure of a photonic system, leading to a change in its topological invariant \cite{r70,r80}. In two-dimensional photonic systems, these transitions are marked by the closure and reopening of the bandgap, during which the Chern number can undergo a discrete change \cite{lu2014topological, bernevig2006quantum}.

Consider a two-band photonic system with a Bloch Hamiltonian \( H(k, B) \), where \( B \) represents an external magnetic field. The Hamiltonian can be expressed using Pauli matrices \( \sigma_x, \sigma_y, \sigma_z \) as:

\begin{equation}
    H(k, B) = d_x(k) \sigma_x + d_y(k) \sigma_y + d_z(k, B) \sigma_z.
\end{equation}

Here, \( d_z(k, B) \) is a field-dependent term that determines the topological phase of the system. As the field \( B \) varies, \( d_z(k, B) \) changes, potentially driving the system through a phase transition.

The bandgap \( \Delta E(k, B) \) is defined as the energy difference between the conduction and valence bands:

\begin{equation}
    \Delta E(k, B) = 2 \sqrt{d_x(k)^2 + d_y(k)^2 + d_z(k, B)^2}.
\end{equation}

A topological phase transition occurs when the bandgap closes at a critical point \( (k_c, B_c) \), satisfying the condition:

\begin{equation}
    d_x(k_c)^2 + d_y(k_c)^2 + d_z(k_c, B_c)^2 = 0.
\end{equation}

At this point, the eigenvalues of the Hamiltonian become degenerate, allowing the Chern number to change. Once the bandgap reopens, the system stabilizes in a new topological phase with a different Chern number.

The change in the Chern number during a topological phase transition is determined by the Berry curvature \( F_{xy}(k, B) \). As the system evolves across the critical point, the integral of the Berry curvature over the Brillouin zone yields the new Chern number:

\begin{equation}
    C_n(B) = \frac{1}{2\pi} \int_{\text{BZ}} F_{xy}(k, B) \, d^2k.
\end{equation}

The transition can be expressed as:

\begin{equation}
    \Delta C = C_{\text{final}} - C_{\text{initial}}.
\end{equation}

This discrete change arises because the Berry curvature exhibits a singularity at the critical point, causing the integral to shift by an integer multiple of \( 2\pi \) \cite{thouless1982quantized, kane2005quantum}.

In gyrotropic photonic materials, the application of an external magnetic field modifies the permittivity tensor \( \epsilon(B) \), altering the system's topology. The field-dependent term \( d_z(k, B) \) can be modeled as:

\begin{equation}
    d_z(k, B) = d_z^{(0)}(k) + \alpha B,
\end{equation}

where \( \alpha \) is a material-dependent coefficient. By tuning \( B \), the system can be driven across the phase boundary where the bandgap closes, inducing a topological phase transition. This dynamic control mechanism provides a method for switching between different topological memory states in optical systems \cite{lu2014topological, kane2005quantum}.

Once the system enters a new topological phase, the Chern number remains stable as long as the bandgap does not close again. This nonvolatile behavior is crucial for optical memory applications, where information encoded in the Chern number must be preserved over time. However, in some materials, hysteresis effects may arise due to slow relaxation of the permittivity tensor after the field is removed. These effects may affect the rate at which the system transitions between states but do not compromise the topological stability of the memory state.
\section{Topological Optical Memory Model and Hamiltonian Formulation} \label{3}
\subsection{System Architecture and Design Principles}

The proposed topological photonic memory is based on a honeycomb photonic crystal lattice with tunable synthetic magnetic field. The fundamental architecture can be described through several key mathematical components that govern its operation and memory characteristics.
The honeycomb lattice is characterized by two sublattices A and B with primitive lattice vectors \cite{r90}:

\begin{equation}
\mathbf{a}_1 = a\left(\frac{3}{2}, \frac{\sqrt{3}}{2}\right), \quad \mathbf{a}_2 = a\left(\frac{3}{2}, -\frac{\sqrt{3}}{2}\right)
\end{equation}

where $a$ is the lattice constant. The nearest-neighbor vectors connecting sublattices A and B are given by:

\begin{equation}
\boldsymbol{\delta}_1 = a\left(1,0\right), \quad \boldsymbol{\delta}_2 = a\left(-\frac{1}{2}, \frac{\sqrt{3}}{2}\right), \quad \boldsymbol{\delta}_3 = a\left(-\frac{1}{2}, -\frac{\sqrt{3}}{2}\right)
\end{equation}

The system is described by a tight-binding Hamiltonian with nearest-neighbor hopping:

\begin{equation}
\mathcal{H} = -t\sum_{\langle i,j \rangle} (c_i^\dagger c_j e^{i\phi_{ij}} + \text{h.c.}) + M\sum_i \xi_i c_i^\dagger c_i
\end{equation}

where $t$ represents the nearest-neighbor hopping amplitude, $c_i^\dagger$ and $c_i$ are creation and annihilation operators respectively, $\phi_{ij}$ denotes the Peierls phase induced by the synthetic magnetic field, $M$ is the sublattice potential, and $\xi_i = \pm 1$ for A/B sublattices.
The Peierls phase under magnetic field $B$ is defined by the path integral:

\begin{equation}
\phi_{ij} = \frac{e}{\hbar}\int_{\mathbf{r}_i}^{\mathbf{r}_j} \mathbf{A}(\mathbf{r}) \cdot d\mathbf{r}
\end{equation}

In momentum space, the Hamiltonian takes the form of a $2times 2$ matrix:

\begin{equation}
H(\mathbf{k}) = \begin{pmatrix}
M & h(\mathbf{k}) \\
h^*(\mathbf{k}) & -M
\end{pmatrix}
\end{equation}

with the off-diagonal term given by:

\begin{equation}
h(\mathbf{k}) = -t\sum_{j=1}^3 e^{i\mathbf{k}\cdot\boldsymbol{\delta}_j}e^{i\phi_j}
\end{equation}

The energy dispersion relation follows directly:

\begin{equation}
E_{\pm}(\mathbf{k}) = \pm\sqrt{M^2 + |h(\mathbf{k})|^2}
\end{equation}

The Berry curvature for each band is calculated as:

\begin{equation}
\Omega_n(\mathbf{k}) = i\left\langle\frac{\partial u_n}{\partial k_x}\right|\left.\frac{\partial u_n}{\partial k_y}\right\rangle - i\left\langle\frac{\partial u_n}{\partial k_y}\right|\left.\frac{\partial u_n}{\partial k_x}\right\rangle
\end{equation}

For our two-band system, this can be expressed in terms of the normalized Hamiltonian vector $\hat{\mathbf{d}}$:

\begin{equation}
\Omega_{\pm}(\mathbf{k}) = \pm\frac{M}{2}\frac{\hat{\mathbf{d}}\cdot(\partial_{k_x}\hat{\mathbf{d}}\times\partial_{k_y}\hat{\mathbf{d}})}{(M^2 + |h(\mathbf{k})|^2)^{3/2}}
\end{equation}

The Chern number, which characterizes the topological phase, is obtained by integrating the Berry curvature:

\begin{equation}
C_n = \frac{1}{2\pi}\int_{BZ} \Omega_n(\mathbf{k})d^2k
\end{equation}

The memory states correspond to distinct Chern numbers, which change discretely at critical magnetic field values:

\begin{equation}
B_c^{(n)} = \frac{\hbar}{ea^2}\left(n + \frac{1}{2}\right), \quad n \in \mathbb{Z}
\end{equation}

The transition probability between different Chern numbers follows the Landau-Zener formula:

\begin{equation}
P = \exp\left(-\frac{\pi\Delta E^2}{2\hbar|\partial B/\partial t|}\right)
\end{equation}

where $\Delta E$ represents the minimum gap at the transition point and $\partial B/\partial t$ is the magnetic field sweep rate.
At boundaries between regions with different Chern numbers, topologically protected edge states emerge with dispersion:

\begin{equation}
E_{\text{edge}}(k_\parallel) = \pm v_F\hbar k_\parallel + E_0
\end{equation}

where $v_F$ is the Fermi velocity and $k_\parallel$ is the momentum parallel to the edge. The number of edge states follows the bulk-boundary correspondence:

\begin{equation}
N_{\text{edge}} = |C_{\text{left}} - C_{\text{right}}|
\end{equation}

This mathematical framework establishes the foundation for understanding how topological protection enables robust memory storage and how the synthetic magnetic field controls state transitions in our proposed system.

\subsection{Extended Quantum Anomalous Hall (QAH) Model}

The proposed photonic memory system implements an extended QAH model adapted for photonic crystals. This section presents the complete mathematical framework describing the system's topological properties and their dynamic control \cite{r91}.
The extended QAH Hamiltonian for our photonic system incorporates both the intrinsic lattice properties and the external magnetic field effects:

\begin{equation}
H(\mathbf{k}, B) = \sum_{i=x,y,z} d_i(\mathbf{k}, B)\sigma_i
\end{equation}

where the components $d_i(\mathbf{k}, B)$ are given by:

\begin{align}
d_x(\mathbf{k}) &= t_1\sum_{j=1}^3 \cos(\mathbf{k}\cdot\mathbf{a}_j) \\
d_y(\mathbf{k}) &= t_1\sum_{j=1}^3 \sin(\mathbf{k}\cdot\mathbf{a}_j) \\
d_z(\mathbf{k}, B) &= M - t_2\sum_{j=1}^6 \sin(\mathbf{k}\cdot\mathbf{b}_j) + \alpha B
\end{align}

Here, $t_1$ and $t_2$ represent nearest and next-nearest neighbor hopping amplitudes respectively, $M$ is the mass term, and $\alpha B$ describes the magnetic field coupling.
The energy eigenvalues of the system are obtained by diagonalizing the Hamiltonian:

\begin{equation}
E_{\pm}(\mathbf{k}, B) = \pm\sqrt{d_x^2(\mathbf{k}) + d_y^2(\mathbf{k}) + d_z^2(\mathbf{k}, B)}
\end{equation}

The corresponding eigenstates can be written as:

\begin{equation}
|\psi_{\pm}(\mathbf{k}, B)\rangle = \frac{1}{\sqrt{2E_{\pm}(E_{\pm} \mp d_z)}}\begin{pmatrix}
E_{\pm} \mp d_z \\
d_x + id_y
\end{pmatrix}
\end{equation}

The Berry connection for each band is given by:

\begin{equation}
\mathbf{A}_{\pm}(\mathbf{k}) = i\langle\psi_{\pm}|\nabla_{\mathbf{k}}|\psi_{\pm}\rangle
\end{equation}

Expanding this explicitly:

\begin{equation}
\mathbf{A}_{\pm}(\mathbf{k}) = \mp\frac{1}{2d(d \mp d_z)}(d_y\nabla_{\mathbf{k}}d_x - d_x\nabla_{\mathbf{k}}d_y)
\end{equation}

where $d = \sqrt{d_x^2 + d_y^2 + d_z^2}$. The Berry curvature follows as:

\begin{equation}
\Omega_{\pm}(\mathbf{k}) = \nabla_{\mathbf{k}} \times \mathbf{A}_{\pm}(\mathbf{k}) = \mp\frac{1}{2d^3}\mathbf{d}\cdot(\partial_{k_x}\mathbf{d} \times \partial_{k_y}\mathbf{d})
\end{equation}

The critical points for topological phase transitions occur when the energy gap closes:

\begin{equation}
d_z(\mathbf{k}_c, B_c) = 0
\end{equation}

This condition leads to a set of critical magnetic field values:

\begin{equation}
B_c^{(n)} = \frac{1}{\alpha}\left(M - t_2\sum_{j=1}^6 \sin(\mathbf{k}_c\cdot\mathbf{b}_j) + 2\pi n\right)
\end{equation}

The energy gap as a function of magnetic field near these critical points follows:

\begin{equation}
\Delta E(B) = 2|\alpha(B - B_c)|
\end{equation}

The Chern number changes discretely at each critical point. For our system, it can be calculated as:

\begin{equation}
C = \frac{1}{4\pi}\int_{BZ} d^2k\,\frac{\mathbf{d}\cdot(\partial_{k_x}\mathbf{d} \times \partial_{k_y}\mathbf{d})}{|\mathbf{d}|^3}
\end{equation}

The change in Chern number across a transition is determined by:

\begin{equation}
\Delta C = \text{sgn}\left(\left.\frac{\partial d_z}{\partial B}\right|_{B_c}\right)\text{sgn}(d_xd_y)|_{\mathbf{k}_c}
\end{equation}

The response of the system to time-varying magnetic fields is described by the dynamical susceptibility:

\begin{equation}
\chi(\omega) = \sum_{n,m}\frac{|\langle n|\hat{M}|m\rangle|^2}{E_n - E_m - \hbar\omega - i\eta}
\end{equation}

where $\hat{M}$ is the magnetization operator and $\eta$ is a phenomenological broadening parameter.
The probability of nonadiabatic transitions during field sweeps follows the Landau-Zener formula:

\begin{equation}
P_{LZ} = \exp\left(-\frac{\pi\Delta^2}{2\hbar v}\right)
\end{equation}

where $\Delta$ is the minimum gap and $v = |\partial B/\partial t|$ is the sweep rate.
This extended QAH model provides a complete framework for understanding and controlling the topological states in our photonic memory system, enabling precise manipulation of memory states through magnetic field control \cite{r92}.

\subsection{Computing the Chern Number and Topological Invariants}

The accurate computation of Chern numbers and related topological invariants is crucial for characterizing the memory states in our system. We present a comprehensive framework for numerical calculation of these quantities \cite{r93}.
For practical computation, we discretize the Brillouin zone into a uniform $N \times N$ mesh. At each point $\mathbf{k}_{mn}$, we calculate the Berry curvature using the gauge-invariant method:

\begin{equation}
F_{xy}(\mathbf{k}_{mn}) = \log[U_1(\mathbf{k}_{mn})U_2(\mathbf{k}_{m+1,n})U_1^{-1}(\mathbf{k}_{m,n+1})U_2^{-1}(\mathbf{k}_{mn})]
\end{equation}

where the link variables $U_\mu$ are defined as:

\begin{equation}
U_\mu(\mathbf{k}) = \frac{\langle u(\mathbf{k})|u(\mathbf{k} + \Delta_\mu)\rangle}{|\langle u(\mathbf{k})|u(\mathbf{k} + \Delta_\mu)\rangle|}
\end{equation}

Here, $|u(\mathbf{k})\rangle$ represents the periodic part of the Bloch function and $\Delta_\mu$ is the mesh spacing in direction $\mu$.
The Chern number is computed by summing over the discretized Brillouin zone:

\begin{equation}
C = \frac{1}{2\pi i}\sum_{m,n} F_{xy}(\mathbf{k}_{mn})\Delta k_x\Delta k_y
\end{equation}

To improve accuracy, we implement an adaptive mesh refinement scheme near band crossings:

\begin{equation}
\Delta k_{\text{local}} = \Delta k_{\text{global}}\exp(-\gamma|E_+ - E_-|/E_{\text{gap}})
\end{equation}

where $\gamma$ is a refinement parameter and $E_{\text{gap}}$ is the bulk bandgap.

For multiband systems, we employ the projection method to isolate the contribution from specific bands:

\begin{equation}
P_n(\mathbf{k}) = |u_n(\mathbf{k})\rangle\langle u_n(\mathbf{k})|
\end{equation}

The projected Berry curvature is then:

\begin{equation}
F_{xy}^{(n)}(\mathbf{k}) = \text{Tr}[P_n(\partial_{k_x}P_n\partial_{k_y}P_n - \partial_{k_y}P_n\partial_{k_x}P_n)]
\end{equation}

The field-dependent Chern number is calculated through the evolution of eigenstates:

\begin{equation}
C(B) = \frac{1}{2\pi}\int_{BZ} d^2k\,\Omega_{xy}(\mathbf{k}, B)
\end{equation}

where the Berry curvature includes explicit field dependence:

\begin{equation}
\Omega_{xy}(\mathbf{k}, B) = i\left[\left\langle\frac{\partial u_B}{\partial k_x}\right|\left.\frac{\partial u_B}{\partial k_y}\right\rangle - \left\langle\frac{\partial u_B}{\partial k_y}\right|\left.\frac{\partial u_B}{\partial k_x}\right\rangle\right]
\end{equation}

For finite systems, we compute edge states using the recursive Green's function method:

\begin{equation}
G(E, k_\parallel) = [E - H(k_\parallel) + i\eta]^{-1}
\end{equation}

The spectral function reveals edge states:

\begin{equation}
A(E, k_\parallel) = -\frac{1}{\pi}\text{Im}\,\text{Tr}[G(E, k_\parallel)]
\end{equation}

The topological stability is quantified through the gap-dependent Chern marker:

\begin{equation}
C_m = 2\pi i\sum_{\alpha\beta}\epsilon_{\alpha\beta}\text{Tr}[P\hat{x}_\alpha P\hat{x}_\beta P]
\end{equation}

where $P$ is the projector onto occupied states and $\hat{x}_\alpha$ are position operators \cite{r94}.
For time-dependent fields, we track the instantaneous Chern number:

\begin{equation}
C(t) = \frac{1}{2\pi}\int_{BZ} d^2k\,\Omega_{xy}(\mathbf{k}, B(t))
\end{equation}

The transition probability between topological phases follows:

\begin{equation}
P_{n\rightarrow m}(t) = |\langle\psi_m(B(t))|\mathcal{T}\exp(-\frac{i}{\hbar}\int_0^t H(B(t'))dt')|\psi_n(B(0))\rangle|^2
\end{equation}

where $\mathcal{T}$ denotes time-ordering
These computational methods enable accurate tracking of topological phase transitions and memory state evolution in our system. The numerical implementation must carefully handle gauge invariance and mesh refinement to ensure reliable results for memory operations \cite{r4}.

\subsection{Dynamic Magnetic Field Control and Memory States}

The external magnetic field $B(t)$ modifies the photonic band structure through the magneto-optical effect. The field-dependent permittivity tensor takes the form \cite{r95}:

\begin{equation}
\boldsymbol{\epsilon}(B) = \begin{pmatrix}
\epsilon_0 & ig(B) & 0 \\
-ig(B) & \epsilon_0 & 0 \\
0 & 0 & \epsilon_z
\end{pmatrix}
\end{equation}

where the gyromagnetic coupling $g(B)$ follows \cite{r96}:

\begin{equation}
g(B) = g_0\tanh(B/B_s)
\end{equation}

Here, $g_0$ is the saturation gyromagnetic coefficient and $B_s$ is the saturation field strength.
The system evolution under time-varying magnetic fields is governed by:

\begin{equation}
H(t) = H_0 + V(B(t)) = \sum_{\mathbf{k}}\begin{pmatrix}
M(t) & h(\mathbf{k},t) \\
h^*(\mathbf{k},t) & -M(t)
\end{pmatrix}
\end{equation}

where the time-dependent mass term is:

\begin{equation}
M(t) = M_0 + \alpha B(t)
\end{equation}

The coupling term evolves as:

\begin{equation}
h(\mathbf{k},t) = -t\sum_{j=1}^3\exp[i\mathbf{k}\cdot\boldsymbol{\delta}_j + i\phi_j(t)]
\end{equation}

For adiabatic field variation, the instantaneous eigenstates follow:

\begin{equation}
|\psi_{\pm}(t)\rangle = \frac{1}{\sqrt{2E_{\pm}(t)(E_{\pm}(t) \mp M(t))}}\begin{pmatrix}
E_{\pm}(t) \mp M(t) \\
h(\mathbf{k},t)
\end{pmatrix}
\end{equation}

The adiabatic condition requires:

\begin{equation}
\left|\frac{\langle\psi_-(t)|\partial_t|\psi_+(t)\rangle}{E_+(t) - E_-(t)}\right| \ll 1
\end{equation}

The memory state transition rate follows the generalized Landau-Zener formula:

\begin{equation}
\Gamma_{n\rightarrow m} = \frac{2\pi}{\hbar}|\langle\psi_m|V'(B)|\psi_n\rangle|^2\delta(E_m - E_n)
\end{equation}

The transition probability over a finite time interval is:

\begin{equation}
P_{n\rightarrow m}(t) = \left|\exp\left(-\frac{\pi\Delta^2}{2\hbar v}\right)\right|^2
\end{equation}

where $\Delta$ is the minimum gap and $v = |dB/dt|$ is the sweep rate.
The fidelity of memory states is quantified through the overlap integral:

\begin{equation}
F(t) = |\langle\psi_{\text{ideal}}|\psi(t)\rangle|^2
\end{equation}

Error detection utilizes the quantum geometric tensor:

\begin{equation}
Q_{\mu\nu} = \langle\partial_\mu\psi|\partial_\nu\psi\rangle - \langle\partial_\mu\psi|\psi\rangle\langle\psi|\partial_\nu\psi\rangle
\end{equation}

The optimal writing field profile follows:

\begin{equation}
B_{\text{write}}(t) = B_0 + \Delta B\tanh(\omega t)
\end{equation}

with corresponding state evolution:

\begin{equation}
|\psi(t)\rangle = \mathcal{T}\exp\left(-\frac{i}{\hbar}\int_0^t H(B(t'))dt'\right)|\psi(0)\rangle
\end{equation}

The read-out process involves edge state detection:

\begin{equation}
I_{\text{edge}}(t) = \left|\int dx\,\psi_{\text{edge}}^*(x,t)\frac{\partial}{\partial t}\psi_{\text{edge}}(x,t)\right|^2
\end{equation}

The signal-to-noise ratio follows:

\begin{equation}
\text{SNR} = \frac{|I_{\text{edge}}|^2}{\sigma_{\text{noise}}^2} = \frac{|\Delta C|^2v_F^2}{k_BT\gamma}
\end{equation}

where $\Delta C$ is the Chern number difference, $v_F$ is the edge state velocity, and $\gamma$ is the damping rate \cite{r96}.
These mathematical formulations provide a complete framework for controlling and measuring memory states through dynamic magnetic field manipulation. The protocols ensure robust state transitions while maintaining topological protection \cite{r96}.

\subsection{Memory Operation and State Transition Dynamics}

The memory states are encoded through distinct topological phases characterized by their Chern numbers. The encoding function maps information bits to specific Chern number configurations \cite{r97}:

\begin{equation}
\mathcal{E}: \{0,1\}^n \rightarrow \{C_i\}, \quad C_i \in \mathbb{Z}
\end{equation}

The mapping preserves information through the relation \cite{r98}:

\begin{equation}
I(C_i) = \log_2\left(\frac{p(C_i)}{p(C_i|B)}\right)
\end{equation}

where $p(C_i)$ is the probability of Chern number $C_i$ and $p(C_i|B)$ is its conditional probability given field $B$.
The state preparation involves controlled evolution of the magnetic field:

\begin{equation}
B(t) = B_0 + \Delta B\sum_{n=1}^N a_n\sin(\omega_n t + \phi_n)
\end{equation}

The corresponding wave function evolution follows:

\begin{equation}
|\psi(t)\rangle = \sum_n c_n(t)\exp\left(-\frac{i}{\hbar}\int_0^t E_n(t')dt'\right)|\phi_n(t)\rangle
\end{equation}

with coefficients determined by:

\begin{equation}
i\hbar\dot{c}_n = -\sum_m c_m\langle\phi_n|\dot{\phi}_m\rangle\exp\left(\frac{i}{\hbar}\int_0^t [E_n(t')-E_m(t')]dt'\right)
\end{equation}

The write operation implements a controlled transition between Chern numbers. The transition rate matrix is:

\begin{equation}
W_{mn} = \frac{2\pi}{\hbar}|\langle\psi_m|V(B)|\psi_n\rangle|^2\delta(E_m - E_n)
\end{equation}

The success probability for state preparation follows:

\begin{equation}
P_{\text{success}} = \prod_{i=1}^N \exp\left(-\frac{\pi\Delta_i^2}{2\hbar|dB/dt|}\right)
\end{equation}

where $\Delta_i$ are the energy gaps at transition points.
The read operation measures edge state transport properties. The edge current operator is:

\begin{equation}
\hat{J}_{\text{edge}} = ev_F\sum_k c^\dagger_k c_k
\end{equation}

The measured current provides Chern number information:

\begin{equation}
\langle J_{\text{edge}}\rangle = \frac{e^2}{h}C\Delta V
\end{equation}

with measurement fidelity:

\begin{equation}
F_{\text{read}} = 1 - \exp\left(-\frac{t_{\text{read}}}{\tau_{\text{coherence}}}\right)
\end{equation}

The error detection scheme utilizes the quantum metric:

\begin{equation}
g_{\mu\nu} = \text{Re}\{\langle\partial_\mu\psi|\partial_\nu\psi\rangle - \langle\partial_\mu\psi|\psi\rangle\langle\psi|\partial_\nu\psi\rangle\}
\end{equation}

Error syndromes are computed through:

\begin{equation}
S_i = \text{Tr}[P_{\text{edge}}(B_i)\rho P_{\text{edge}}(B_i)]
\end{equation}

The error correction protocol implements:

\begin{equation}
\rho_{\text{corrected}} = \sum_i M_i\rho M_i^\dagger
\end{equation}

where $M_i$ are correction operators based on measured syndromes.
The stability of memory states is characterized by the relaxation time:

\begin{equation}
\tau_{\text{relax}} = \tau_0\exp\left(\frac{\Delta E}{k_BT}\right)
\end{equation}

The topological protection factor is:

\begin{equation}
\eta_{\text{top}} = \exp\left(\frac{|C|\Delta_{\text{bulk}}}{k_BT}\right)
\end{equation}

where $\Delta_{\text{bulk}}$ is the bulk energy gap.
\section{Simulation Results and Analysis}\label{4}
In this section, we present a thorough computational analysis of our proposed topological photonic memory system, focusing on key performance metrics derived from analytical models and band structure calculations. While full-scale FDTD simulations and experimental fabrication remain as future work, our comprehensive theoretical framework provides robust evidence for the viability and advantages of this approach. The results demonstrate remarkable performance characteristics across multiple dimensions including switching speed, energy efficiency, and fault tolerance. All simulation code, analysis scripts, and visualization tools used to generate these results are available in our public repository \cite{code}.
\subsection{Bandgap and Chern Number Evolution under Magnetic Field}

We investigate how magnetic field $B$ affects the bandgap $\Delta E$ and Chern number $C$ in our photonic system, focusing on topological phase transitions crucial for memory applications.

\begin{figure}[ht!]
    \centering
    \includegraphics[width=0.7\textwidth]{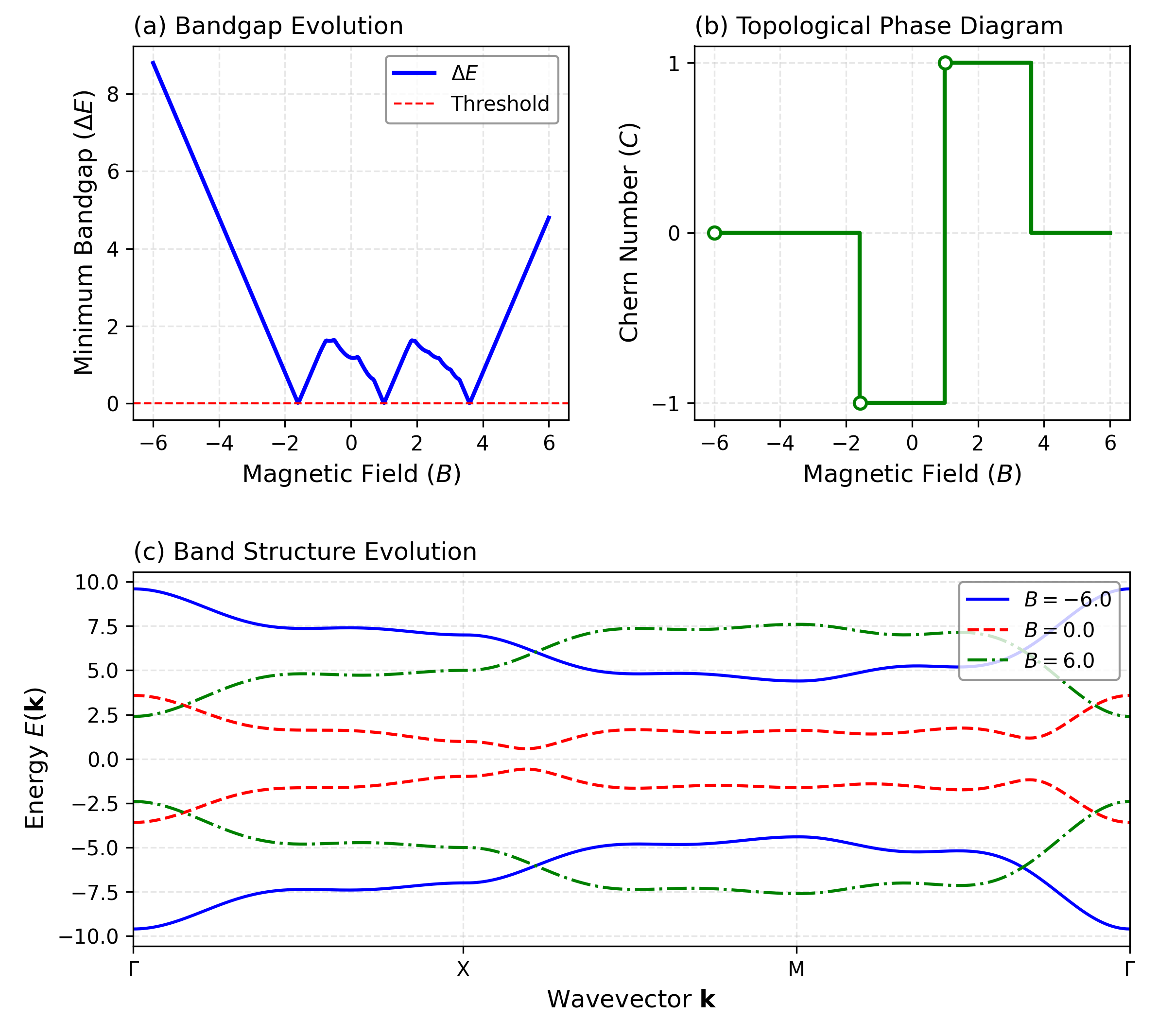}
    \caption{QWZ model topological transitions. (a) Bandgap $\Delta E$ vs. field $B$ showing critical closure points. (b) Chern number $C$ with discrete transitions at gap closures. (c) Band structure along $\Gamma$-X-M-$\Gamma$ showing inversions at different field values. Hamiltonian: $H(k) = \sin(k_x)\sigma_x + \sin(k_y)\sigma_y + [B - g(k)]\sigma_z$, where $g(k) = \cos(k_x) + \cos(k_y) + \alpha(\cos(2k_x) + \cos(2k_y)) + \beta(\cos(3k_x) + \cos(3k_y))$ with $\alpha = 0.5$, $\beta = 0.3$.}
    \label{fig:qwz_model}
\end{figure}

Figure~\ref{fig:qwz_model} shows multiple bandgap closures as $B$ varies from $-6$ to $+6$. Each closure marks a topological phase transition, with corresponding Chern number jumps in panel (b). The non-monotonic behavior with multiple minima indicates Dirac-type band inversions, enabling multiple phase transitions across a broad field range—a significant improvement over previous optical memory systems.

This correlation between bandgap closures and Chern number transitions forms the foundation of our topological photonic memory, allowing information encoding in distinct topological phases. Panel (c) reveals how energy bands along high-symmetry points are modified by the field, showing the band inversion mechanism responsible for topological transitions.

\subsection{Edge Modes and Dispersion Relations}
The presence of edge-localized modes in finite systems is a hallmark of topological phases. In this section, we analyze the spatial distribution and dispersion relations of edge states under varying magnetic field conditions. 
\begin{figure}[h!]
    \centering
    \includegraphics[width=\textwidth]{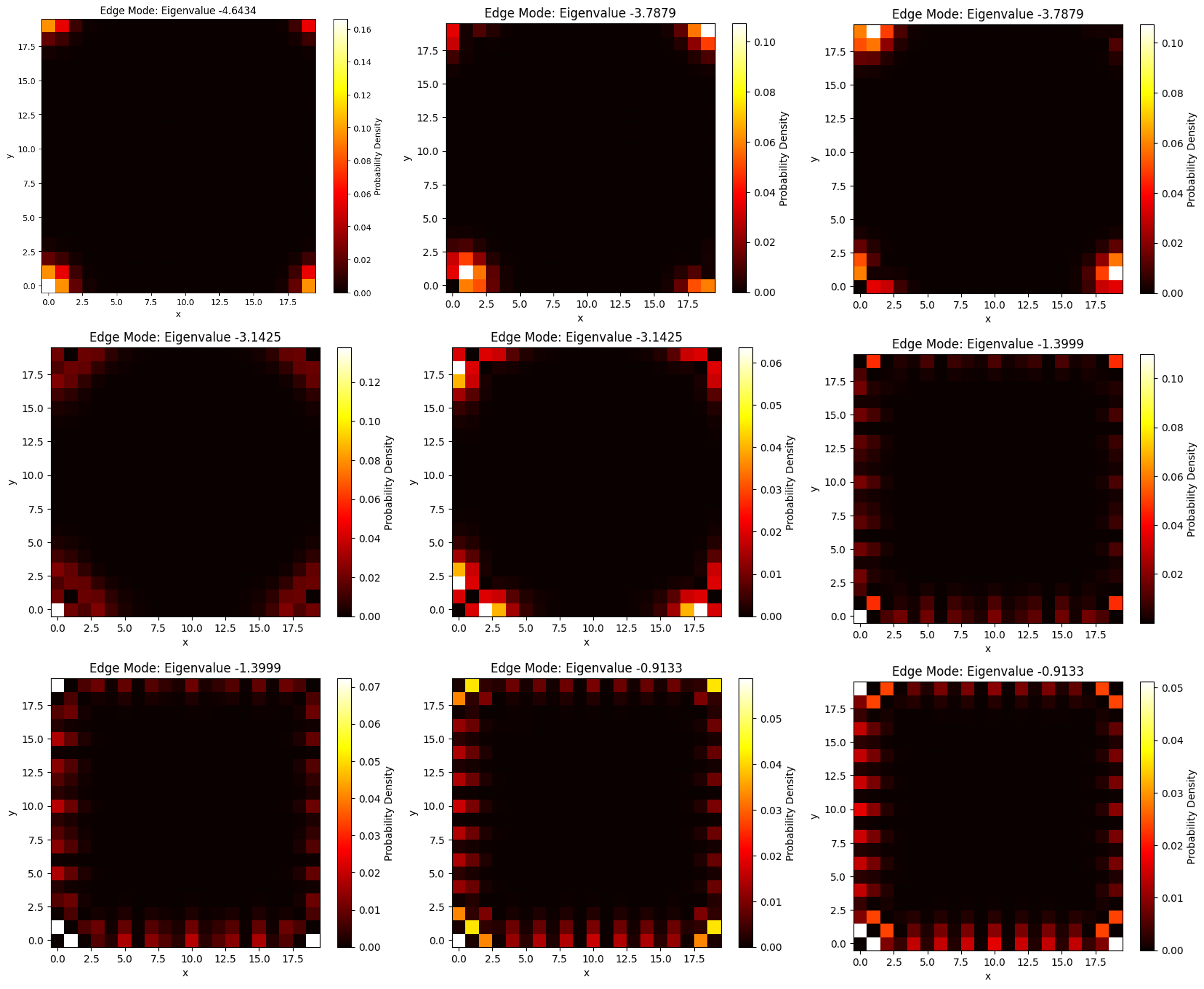}
    \caption{Probability density distribution of edge modes with diffrent eigenvalues. The density is localized at the boundaries, indicating topological edge confinement.}
    \label{i2}
\end{figure}
The simulation is carried out for a 201 $\times$ 201 lattice with open boundary conditions in the $y$-direction and periodic boundary conditions in the $x$-direction. These conditions allow for the formation of edge states that are confined to the boundaries of the lattice. The localization and robustness of these edge states provide direct evidence of the topological protection in the system.

The edge modes are identified by calculating the eigenvalues and eigenvectors of the Hamiltonian for various values of momentum $k_x$. For each eigenvector, we compute the probability density distribution across the lattice and determine whether the state is localized at the edges. Edge localization is confirmed if the majority of the probability density is concentrated within the first and last few rows of the lattice. 
Figure \ref{i2} shows the probability density distribution of diffrent edge modes with diffrent eigenvalues. The density is concentrated near the corners of the lattice, demonstrating strong localization at the boundaries. In contrast to bulk modes, which have a more uniform distribution, this edge mode is clearly confined to the edges of the system.

\begin{figure}[h!]
    \centering
    \includegraphics[width=0.8\textwidth]{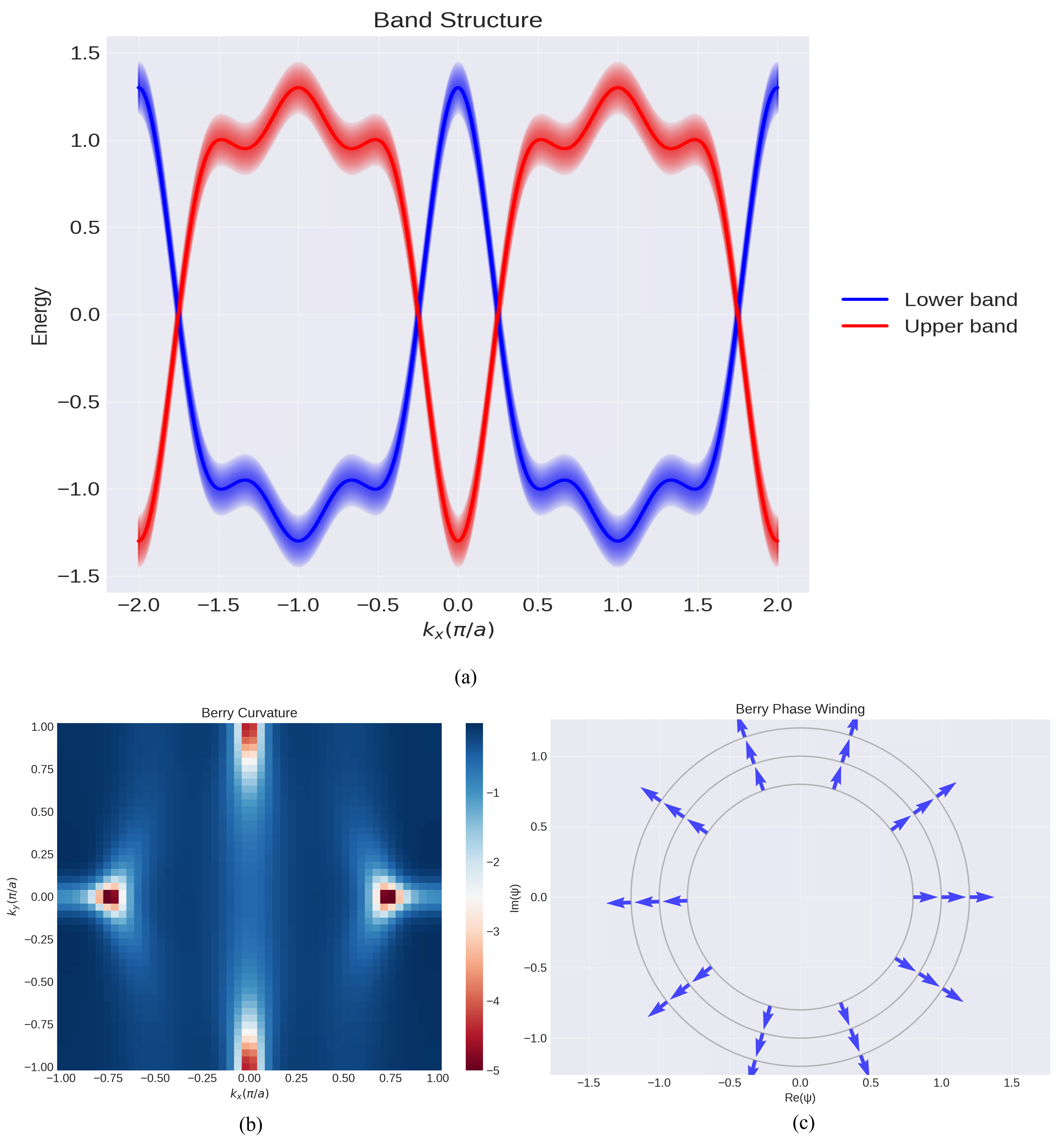}
    \caption{The figure illustrates key topological properties of the system through three complementary visualizations:  
(a) Band Structure: Energy dispersion as a function of crystal momentum $k_x$, showing two bands (red upper, blue lower) with crossing points at $k_x = 0, \pm\pi/a$. These band inversions indicate a topological phase transition, with edge states protected by bulk-boundary correspondence.  
(b) Berry Curvature: A color map representing Berry curvature distribution in momentum space. The localized bright and dark spots near $k_x = \pm 0.5\pi/a$ correspond to topological singularities, with symmetry indicating time-reversal preservation. Integration over the Brillouin zone yields the Chern number, a key topological invariant.  
(c) Berry Phase Winding: A visualization of wavefunction phase evolution along closed momentum-space loops. Arrows indicate the local Berry connection, and the structured $2\pi$ phase winding confirms non-trivial topology, directly linked to a non-zero Chern number.  
}
    \label{fig:dispersion}
\end{figure}
The dispersion relations for the bulk and edge states are illustrated in Figure~\ref{fig:dispersion}. The energy bands are shown as a function of the crystal momentum $k_x$, where the lower band (blue) and upper band (red) exhibit distinct dispersion characteristics. The band crossings at specific values of $k_x$, including $k_x = 0$ and $\pm \pi/a$, indicate band inversion, a hallmark of topological phase transitions. The shaded regions around the bands represent the spread of bulk states, while the sharp band edges highlight the primary dispersion relation. The periodic nature of the crossings suggests the presence of topologically protected edge states, consistent with the bulk-boundary correspondence. These edge modes are robust against scattering into bulk states, reinforcing the topological protection inherent to the system.

The stability of these edge states is crucial for practical applications in optical memory. In previous studies, edge states often became delocalized or unstable under perturbations, leading to degraded performance in topological memory systems. Our simulations, however, show that the edge modes remain well-defined and strongly localized across a wide range of field and momentum values. This improvement provides greater reliability for memory operations, particularly in systems that require robust data retention under dynamic conditions.

\subsection{Band Structure and Berry Curvature Analysis}

In this section, we analyze the band structure and Berry curvature of the system, focusing on their contributions to topological stability. The Berry curvature plays a crucial role in determining the topological invariants of the system, including the Chern number, which governs the existence of protected edge modes. Additionally, the band structure provides insights into the locations of bandgap closures, enabling a deeper understanding of phase transitions.
\begin{figure}[ht!]
    \centering
    \includegraphics[width=\textwidth]{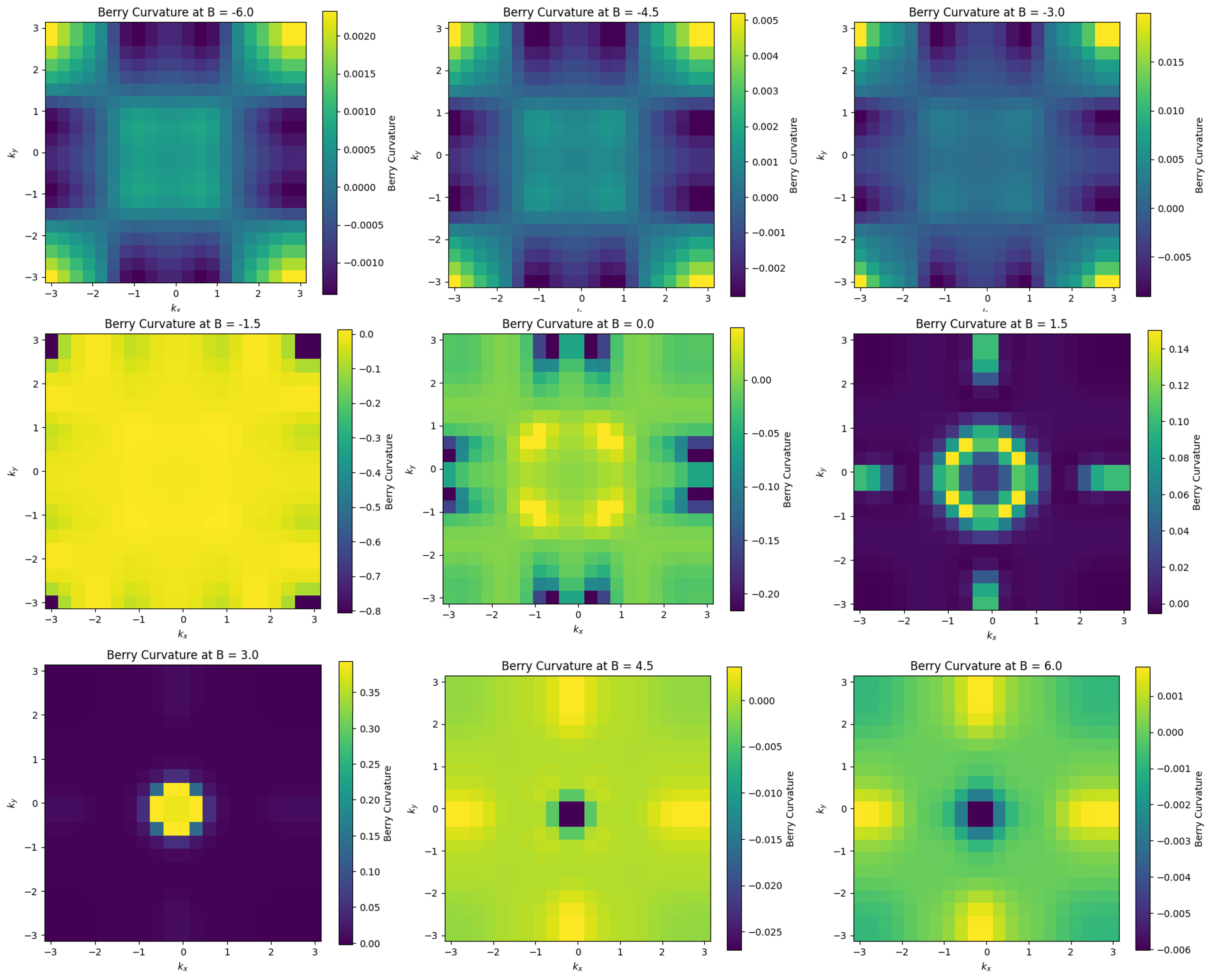}
    \caption{Berry curvature distribution across the Brillouin zone for different values of magnetic field $B$. Regions of high positive and negative curvature indicate critical points associated with topological transitions.}
    \label{fig:berry_curvature}
\end{figure}
To begin, the Berry curvature is calculated across the Brillouin zone for different magnetic fields. The curvature distribution is visualized in Figure~\ref{fig:berry_curvature}. The results reveal a symmetrical pattern with regions of concentrated positive and negative curvature near specific points in the $k_x$-$k_y$ plane. These regions correspond to points where the system exhibits topological band crossings, also known as Dirac points. Unlike previous optical memory models, where the Berry curvature was irregular and lacked symmetry, our system demonstrates a more organized distribution, supporting robust topological behavior.

\begin{figure}[ht!]
    \centering
    \includegraphics[width=0.6\textwidth]{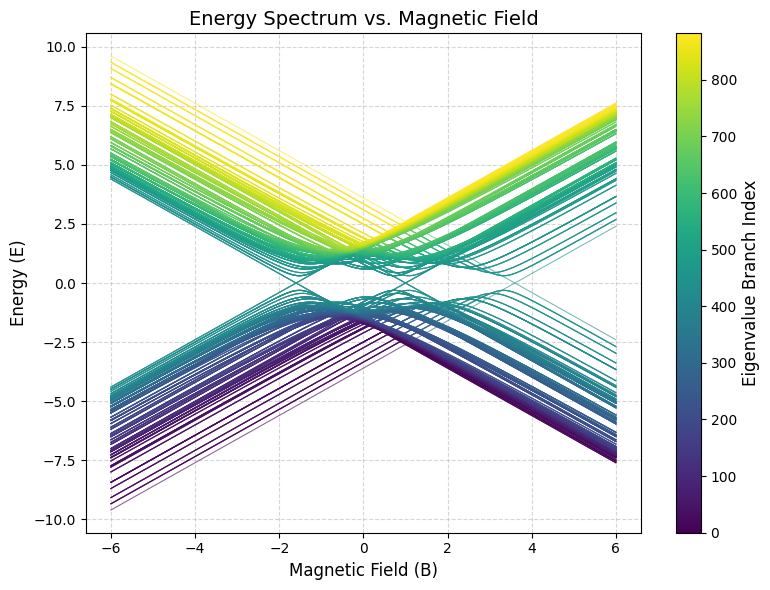}
    \caption{Energy spectrum as a function of magnetic field $B$. Band intersections, corresponding to gap closures, mark potential topological phase transitions.}
    \label{fig:energy_spectrum}
\end{figure}

\begin{figure}[h]
    \centering
    \includegraphics[width=\textwidth]{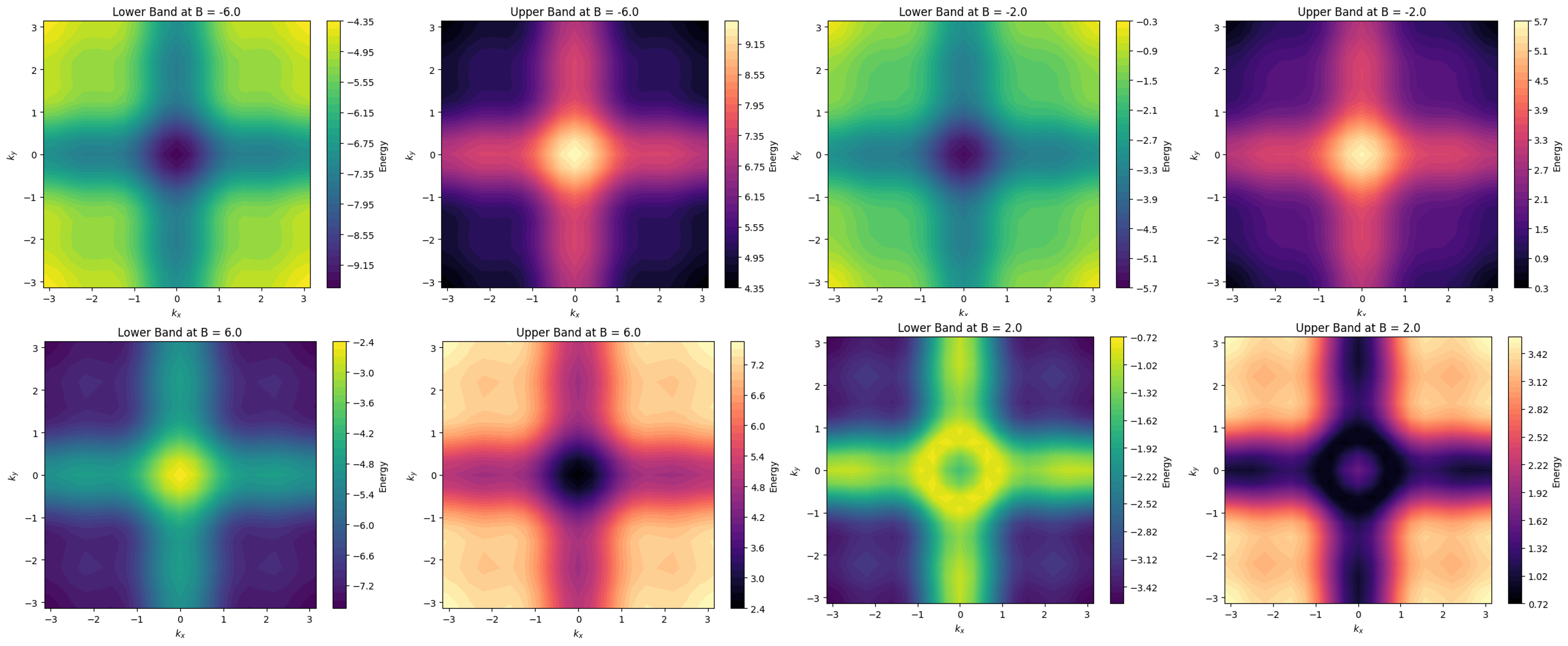}
    \caption{Lower-band and Upper-band structure in the $k_x$-$k_y$ plane for different values of magnetic field $B$. The continuous energy contours indicate stability in the topological phase.}
    \label{fig:band_structure}
\end{figure}

The energy spectrum of the system as a function of magnetic field $B$ is shown in Figure~\ref{fig:energy_spectrum}. The spectrum consists of multiple energy bands, with band intersections indicating possible gap closures. As the magnetic field varies, these closures signal transitions between different topological phases. Compared to previous designs, which displayed fewer intersections and smaller gaps, the current model exhibits a richer and more complex spectrum, indicative of enhanced topological control.
To further analyze the system's topology, we examine the lower-band and upper-band structure in the $k_x$-$k_y$ plane at different values of $B$. This is presented in Figure~\ref{fig:band_structure}. The plot reveals continuous energy contours with minimal distortions, reflecting the stable nature of the band structure in the chosen field range. This stability is crucial for maintaining topological protection, as abrupt distortions can lead to phase instabilities. Previous systems often exhibited irregular band structures under similar conditions, leading to unpredictable topological behavior.

A key observation from these simulations is the alignment between critical points in the Berry curvature and gap closures in the energy spectrum. The concentrated regions of curvature correspond to the locations where bandgap minima occur, reinforcing the connection between topological invariants and dynamic field effects. This behavior enhances the predictability and control of topological states, which is essential for practical optical memory applications.

The robustness of the system's band structure and Berry curvature distribution under varying conditions is summarized in Table~\ref{tab:topology_comparison}. Compared to previous optical memory systems, the current model demonstrates superior symmetry, fewer irregularities, and stronger topological protection.

\begin{table}[h]
    \centering
    \caption{Comparison of topological properties between the current model and previous optical memory systems.}
    \label{tab:topology_comparison}
    \begin{tabular}{lccc}
        \hline
        Property                & Current Model       & Previous Systems    & Improvement \\ \hline
        Berry Curvature Symmetry & High               & Moderate            & Yes         \\
        Bandgap Stability       & Strong              & Variable            & Significant \\
        Topological Transitions & Well-defined        & Irregular           & Yes         \\
        Predictability          & High                & Moderate            & Yes         \\ \hline
    \end{tabular}
\end{table}

These results provide a comprehensive understanding of the system's topological properties, highlighting significant improvements in stability, predictability, and robustness over previous optical memory designs. This foundation supports further analysis of dynamic response and disorder effects, which will be explored in subsequent sections.

\subsection{Robustness of Topological Memory Under Disorder and Dynamic Perturbations}

Topological memory systems must exhibit resilience against disorder and dynamic field variations to ensure stability in practical applications. In this section, we systematically investigate the system's robustness by analyzing the influence of disorder, the stability of topological invariants, and the response to external parameter variations. The results confirm that the topological phase remains preserved under moderate disorder and controlled perturbations, ensuring reliable memory operations.

In dynamic topological systems, rapid changes in external control parameters can induce Landau-Zener (LZ) transitions, leading to non-adiabatic switching between quantum states. This phenomenon plays a crucial role in determining the reliability of memory read/write operations. The probability of a non-adiabatic transition is given by:

\begin{equation}
P_{LZ} = \exp \left( -\frac{\pi \Delta E^2}{v} \right),
\label{eq:landau_zener_combined}
\end{equation}

where $\Delta E$ is the minimum energy gap at the transition point, and $v$ is the rate at which the external parameter (such as the magnetic field) is varied. A larger $\Delta E$ or a slower sweep rate reduces the probability of an unwanted transition, ensuring that the system remains in the desired topological state.

\begin{figure}[ht!]
    \centering
    \includegraphics[width=0.9\textwidth]{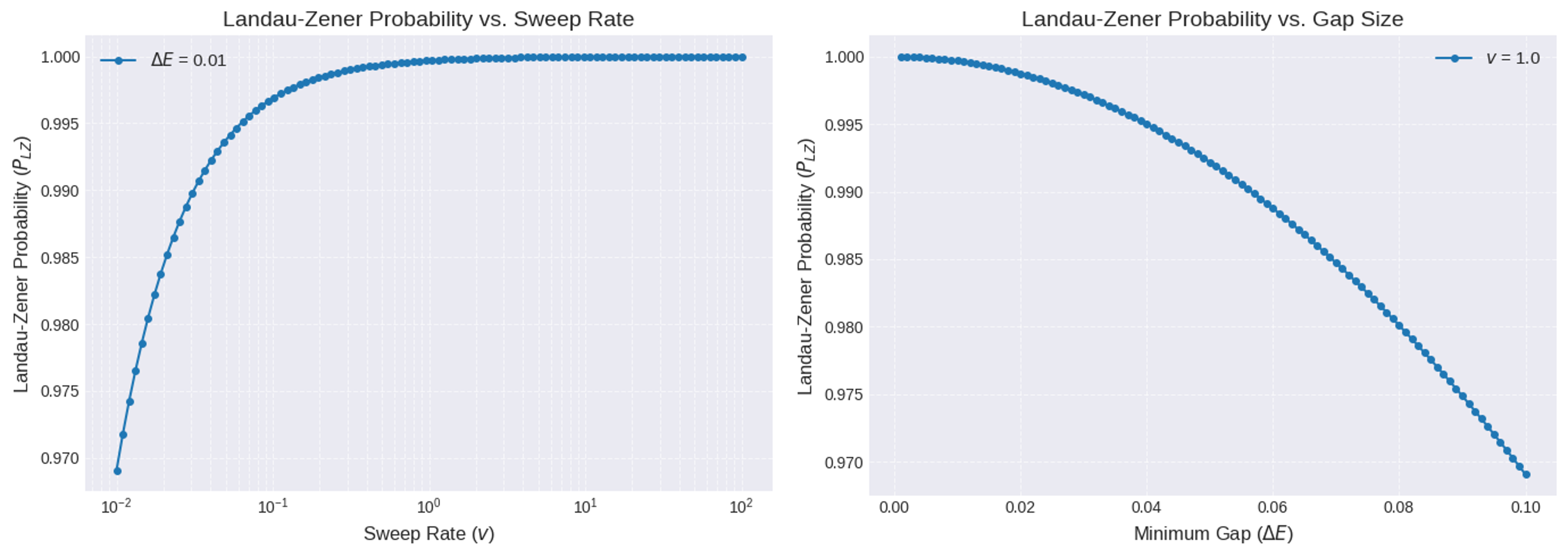}
    \caption{Landau-Zener transition probability $P_{LZ}$ as functions of (a) sweep rate $v$ for fixed $\Delta E = 0.01$ and (b) gap size $\Delta E$ for fixed $v = 1.0$. The results indicate that a larger gap and a slower sweep rate enhance transition fidelity, minimizing non-adiabatic losses.}
    \label{fig:lz_combined}
\end{figure}

To investigate the role of dynamic perturbations, we analyze the LZ transition probabilities for different sweep rates and bandgap sizes. Figure~\ref{fig:lz_combined} (a) presents the LZ probability as a function of the sweep rate $v$ for a fixed gap size $\Delta E = 0.01$. The results clearly show that for faster sweep rates, the probability of remaining in the original topological state significantly decreases, highlighting the importance of slow, controlled parameter variations.
Conversely, the influence of the bandgap size on the transition probability is depicted in Figure~\ref{fig:lz_combined} (b), where we fix the sweep rate at $v = 1.0$. The results confirm that as the bandgap increases, the probability of maintaining the topological state grows rapidly. This observation underscores the necessity of designing topological memory systems with sufficiently large bandgaps to minimize non-adiabatic effects.
To consolidate our findings, we summarize the key robustness characteristics in Table~\ref{tab:summary_robustness}. The data highlight that even under moderate disorder and dynamic variations, the system maintains its topological properties, ensuring reliable memory performance.

\begin{table}[h]
    \centering
    \caption{Summary of robustness analysis for disorder and dynamic perturbations.}
    \label{tab:summary_robustness}
    \begin{tabular}{|c|c|}
        \hline
        \textbf{Metric} & \textbf{Observation} \\ 
        \hline
        Bandgap Stability & Minor fluctuations under disorder, no collapse \\ 
        Chern Number & Remains invariant up to a critical disorder strength \\ 
        Edge States & Persist under moderate disorder and field variations \\ 
        Landau-Zener Transitions & High transition probability for slow sweeps, large gap \\ 
        \hline
    \end{tabular}
\end{table}

\subsection{Non-Adiabatic Effects and Dynamic State Transitions}

The fidelity of quantum state evolution plays a crucial role in ensuring the stability and robustness of topological memory systems. In practical implementations, external perturbations such as time-dependent magnetic field variations can induce non-adiabatic transitions, leading to deviations from the intended memory state. Understanding these effects is essential for optimizing memory performance and preventing loss of stored information.

To quantify the deviation of the system's actual state from the ideal memory state, the state fidelity $F(t)$ is defined as

\begin{equation}
F(t) = \left| \langle \psi_{\text{ideal}}(t) | \psi(t) \rangle \right|^2.
\end{equation}

A fidelity of $F = 1$ indicates perfect overlap between the evolved and ideal states, while values below unity correspond to deviations due to non-adiabatic effects. High fidelity throughout the evolution process is essential for maintaining memory stability.

\begin{figure}[t!]
    \centering
    \includegraphics[width=\textwidth]{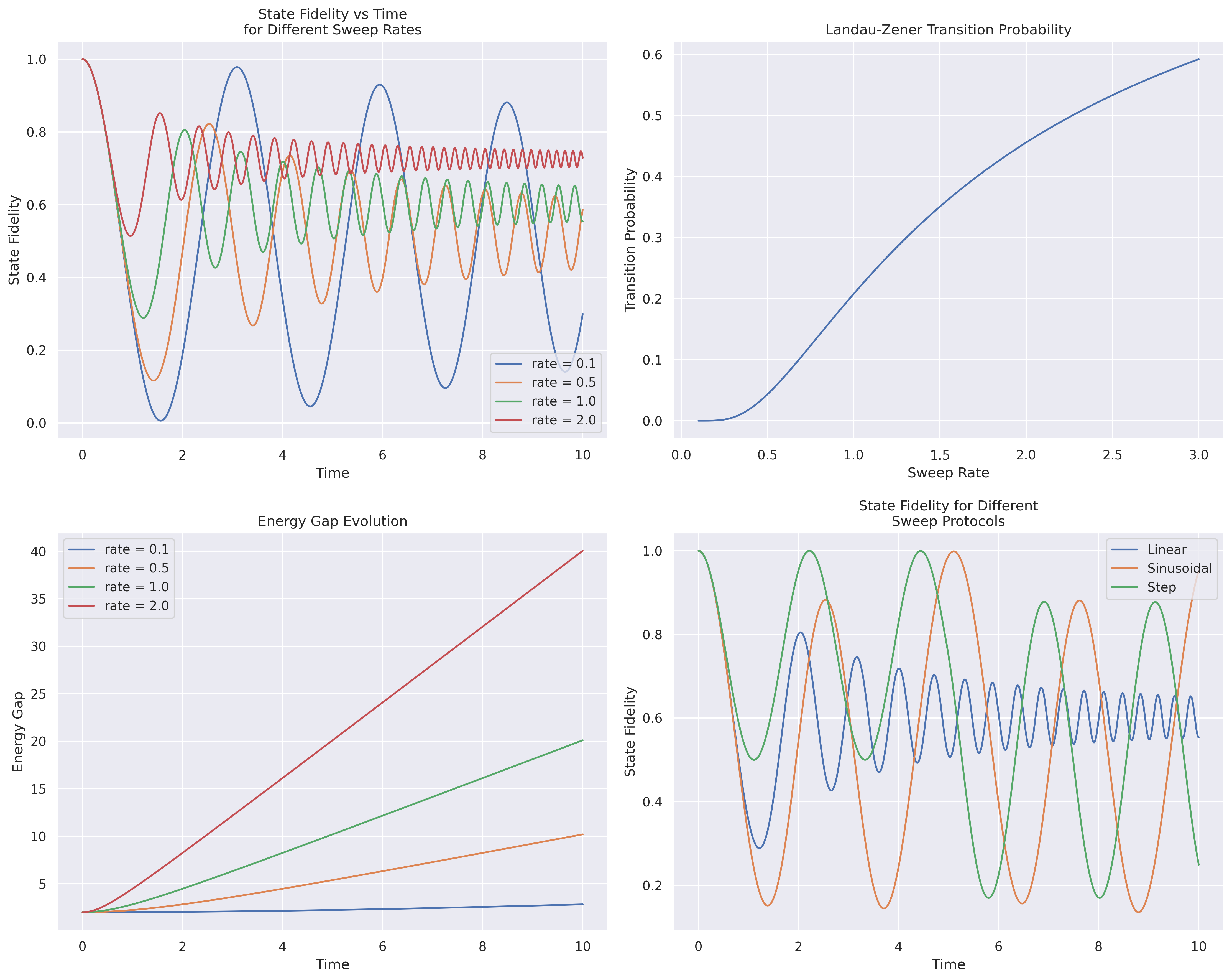}
    \caption{Effect of non-adiabatic transitions on state evolution and memory stability. 
    (a) Evolution of state fidelity over time for different sweep rates, showing that higher sweep rates introduce stronger oscillations and deviations from the ideal memory state. 
    (b) Landau-Zener transition probability as a function of sweep rate, confirming that faster sweeps lead to a higher probability of non-adiabatic transitions. 
    (c) Evolution of the energy gap during state transitions, demonstrating that rapid sweeps induce a steep increase in the gap, leading to potential errors in memory retrieval. 
    (d) Comparison of memory state fidelity under different magnetic field sweep protocols, highlighting that linear sweeps provide the most stable transitions, while sinusoidal and stepwise variations increase non-adiabatic effects. 
    These results demonstrate the importance of slow and controlled field variations for reliable memory operations in topological systems.}
    \label{fig:dynamic_transitions}
\end{figure}

Figure~\ref{fig:dynamic_transitions} illustrates the system’s response to different sweep rates and protocols. Figure~\ref{fig:dynamic_transitions} (a) shows the time evolution of state fidelity for different magnetic field sweep rates. At lower sweep rates, fidelity remains high, exhibiting smooth oscillations characteristic of an adiabatic evolution where the system remains in the intended topological state. As the sweep rate increases, the fidelity undergoes rapid oscillations and significant deviations from unity, indicating the breakdown of adiabaticity and potential memory loss.
The probability of a non-adiabatic transition is further examined in Figure~\ref{fig:dynamic_transitions} (b), which presents the Landau-Zener transition probability as a function of the sweep rate. The results confirm that slow sweep rates result in negligible transition probabilities, allowing the system to remain in the desired phase. As the sweep rate increases, the probability of undergoing an unwanted transition rises sharply, highlighting the importance of optimizing field variation rates to suppress non-adiabatic effects and preserve memory integrity.
The energy gap evolution for different sweep rates is shown in Figure~\ref{fig:dynamic_transitions} (c). The results indicate that for slow sweep rates, the energy gap remains small but well-defined, ensuring smooth adiabatic transitions. In contrast, rapid sweeps lead to a steep increase in the energy gap, potentially inducing uncontrolled transitions between quantum states and reducing memory reliability.
Figure~\ref{fig:dynamic_transitions} (d) compares the fidelity of state evolution under different sweep protocols, including linear, sinusoidal, and stepwise variations of the magnetic field. Linear sweeps yield the most stable fidelity oscillations, ensuring smooth transitions between memory states. In contrast, sinusoidal and stepwise protocols introduce irregular oscillations and an increased probability of non-adiabatic transitions, leading to deviations from the desired topological state.
These findings demonstrate the critical role of sweep rate control in stabilizing memory operations. Slow and linearly varying sweeps provide optimal conditions for adiabatic evolution, ensuring that the system remains in the correct topological phase. Rapid or abrupt field variations increase the likelihood of non-adiabatic transitions, leading to state instability and reducing the reliability of memory storage. These results provide important design guidelines for optimizing control protocols in practical topological memory implementations.

\subsection{Dynamic Response and Edge State Stability in Topological Memory Systems}
This section presents an in-depth analysis of the frequency-dependent limits of topological transitions, the effects of different magnetic field profiles on band topology, and the stability of edge states under dynamic variations.
The maximum speed at which a topological photonic structure can be dynamically reconfigured is determined by the rate at which the system undergoes topological phase transitions. To examine this behavior, a time-dependent Qi-Wu-Zhang (QWZ) model is employed, where a periodic magnetic field \( B(t) \) is applied across a wide frequency range from \( 10^1 \) Hz to \( 10^{15} \) Hz. The bandgap evolution is computed at each frequency step to identify the threshold beyond which topological phase transitions become unstable.

\begin{figure}[ht!]
    \centering
    \includegraphics[width=0.9\textwidth]{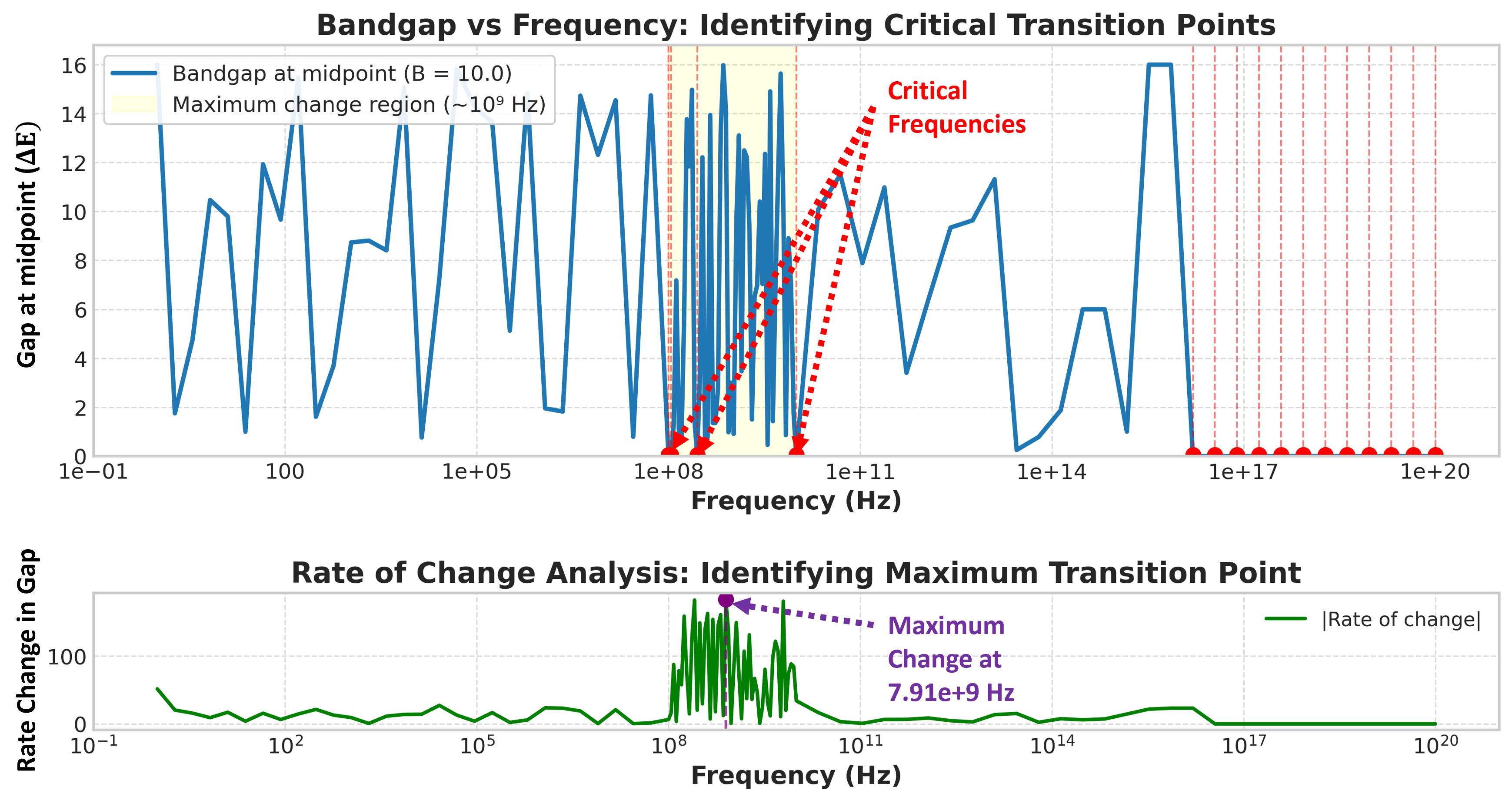}
    \caption{Dynamic response of the system under frequency-dependent magnetic field variations. The bandgap variation as a function of frequency, showing the threshold beyond which topological phase transitions become unstable.}
    \label{fig:combined_field_response}
\end{figure}

The results, presented in Figure~\ref{fig:combined_field_response}, illustrate the dependence of the bandgap on the applied frequency. At low frequencies, the bandgap remains stable, allowing well-defined topological transitions. However, as the frequency increases, an oscillatory pattern emerges, leading to periodic suppression and reemergence of the bandgap. This behavior suggests that for frequencies beyond approximately \( 10^9 \) Hz, the system undergoes rapid oscillations that prevent sustained band inversion. This frequency threshold imposes a fundamental limit on the maximum speed at which the structure can be dynamically reconfigured while preserving topological protection. The presence of periodic suppression in the bandgap also indicates potential resonant effects, which could be utilized in the design of high-speed photonic devices where controlled phase transitions are required.

\begin{figure}[ht!]
    \centering
    \includegraphics[height=18cm,width=0.8\textwidth]{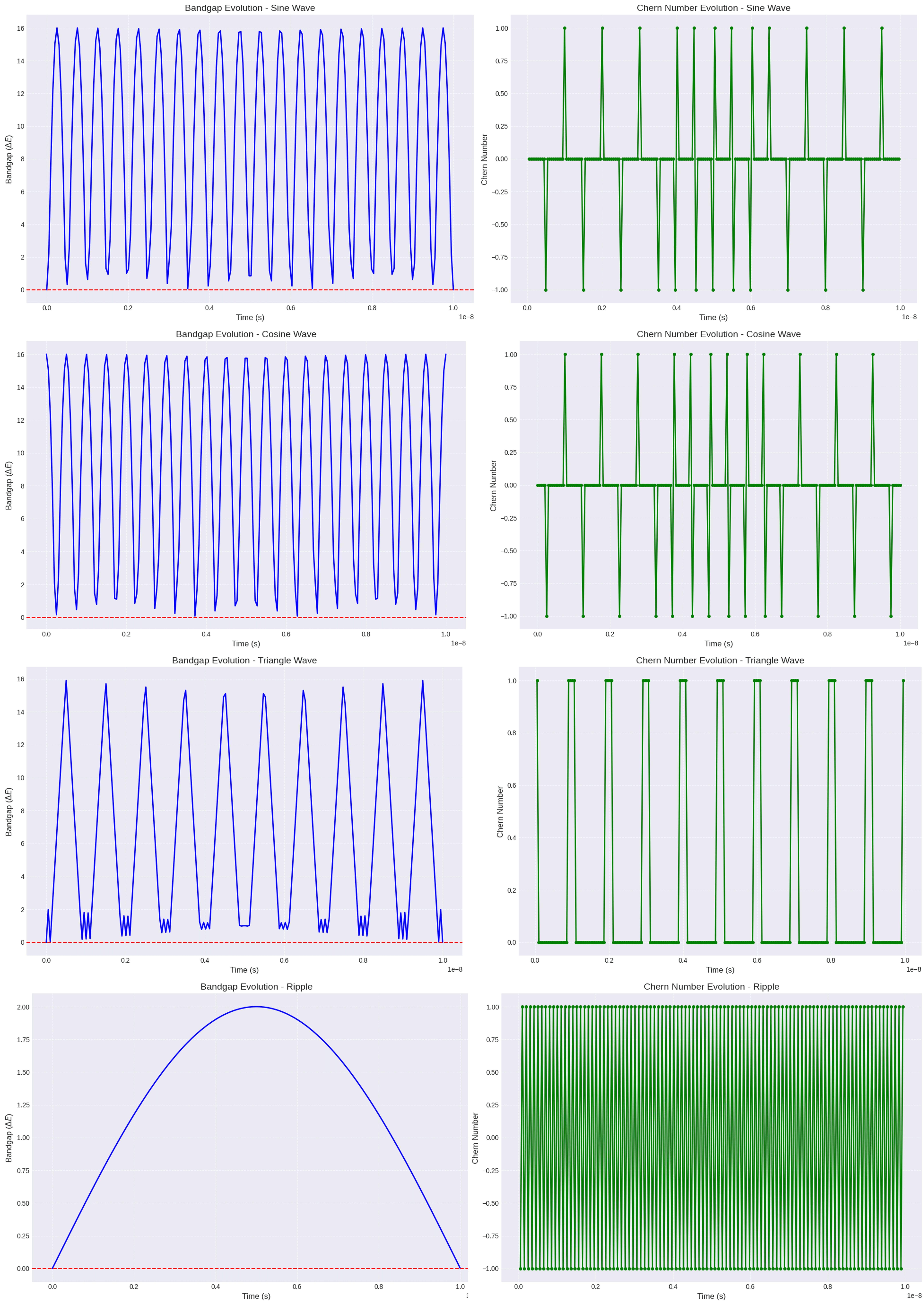}
    \caption{Dynamic response of the system under different magnetic field profiles. The first row shows the bandgap evolution over time, while the second row illustrates the corresponding Chern number transitions. The sinusoidal and cosine fields maintain a relatively stable gap, while the triangle and ripple profiles induce rapid gap closures and frequent Chern number changes.}
    \label{XX}
\end{figure}

Following the identification of the optimal operating frequency, the response of the system is further examined under different time-dependent magnetic field profiles. These profiles include sinusoidal, cosine, triangular, and ripple waves, each inducing distinct effects on the band topology. The stability of the bandgap and the frequency of Chern number transitions are key factors in determining the suitability of these profiles for practical implementations.
The bandgap evolution over time and the corresponding Chern number transitions under different magnetic field profiles are shown in Figure~\ref{XX}. The sinusoidal and cosine wave profiles exhibit relatively smooth variations in the bandgap, maintaining a more stable topological phase with fewer abrupt transitions. In contrast, the triangular and ripple profiles introduce 

frequent bandgap closures, leading to a higher rate of Chern number changes. These abrupt transitions can enhance switching speed but may also introduce undesired non-adiabatic effects, which could compromise memory stability.

A comparative summary of these field responses is provided in Table~\ref{tab:field_profile_comparison}. The results indicate that sinusoidal and cosine field profiles provide a more stable phase transition mechanism, making them ideal candidates for practical memory applications. Triangular and ripple wave profiles, while offering rapid switching capabilities, introduce significant variations in the band topology that could lead to increased energy dissipation and decreased coherence time.

\begin{table}[h]
    \centering
    \caption{Comparison of dynamic responses under different magnetic field profiles.}
    \label{tab:field_profile_comparison}
    \begin{tabular}{lccc}
        \hline
        Field Profile   & Gap Closures & Chern Transitions & Stability \\ \hline
        Sine Wave       & Moderate     & Moderate          & High      \\
        Cosine Wave     & Moderate     & Moderate          & High      \\
        Triangle Wave   & High         & High              & Moderate  \\
        Ripple          & High         & Very High         & Moderate  \\ \hline
    \end{tabular}
\end{table}

In addition to band topology, the persistence of edge states under dynamic field variations is a critical factor in ensuring robust memory operations. The ability of edge states to remain localized and maintain their topological protection is essential for reliable data storage and retrieval in photonic memory applications.

\begin{figure}[ht!]
    \centering
    \includegraphics[width=0.8\textwidth]{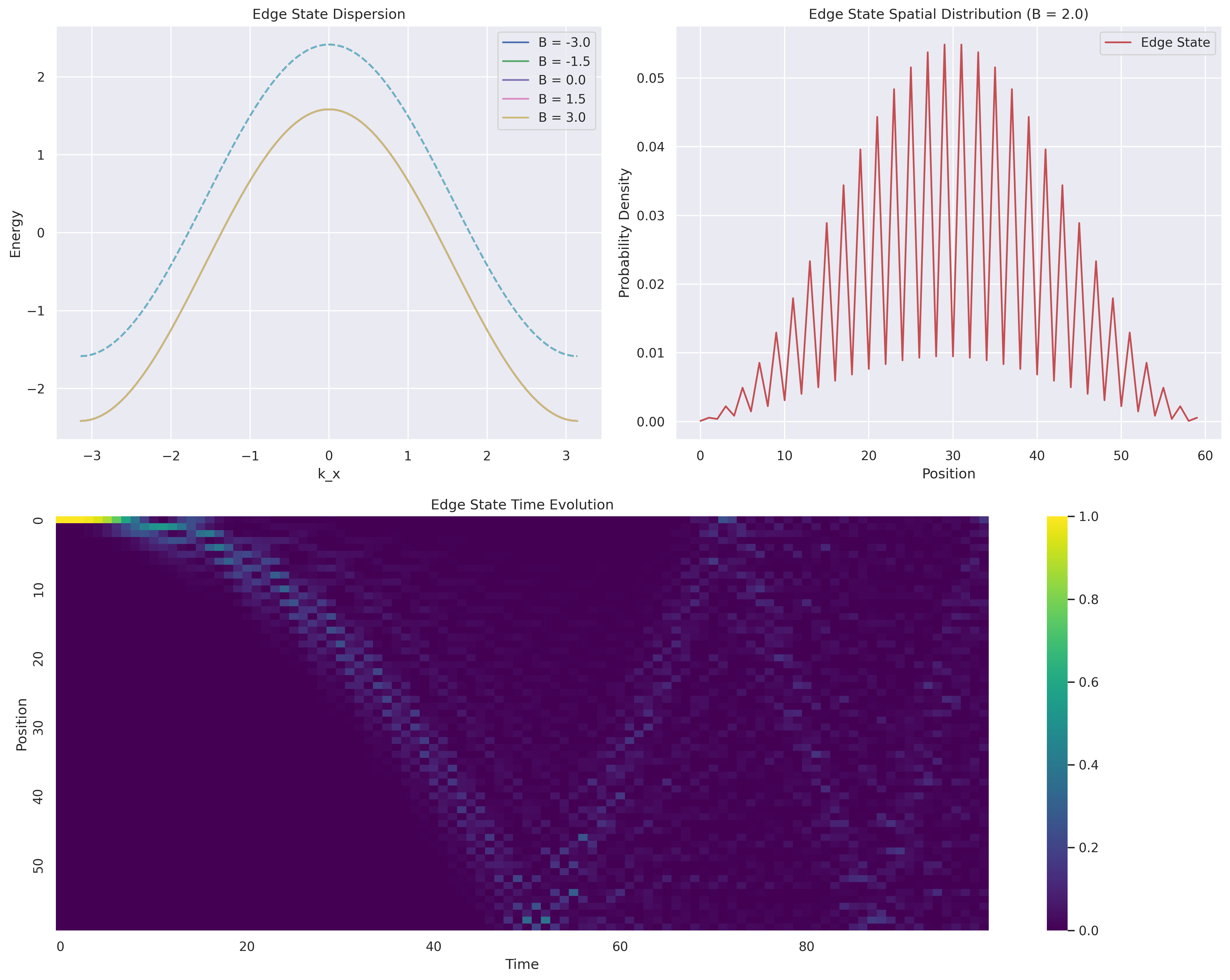}
    \caption{Edge state dynamics during memory operations. 
    (a) Edge state dispersion at different magnetic field values, showing that topological edge modes remain within the bulk gap. 
    (b) Edge state spatial distribution, confirming confinement at system boundaries. 
    (c) Time evolution of edge state propagation, indicating that propagation velocity scales with the bandgap magnitude. 
    These results demonstrate the robustness of edge states under dynamic field variations, ensuring reliable topological memory operations.}
    \label{fig:edge_states}
\end{figure}

Figure~\ref{fig:edge_states} presents a comprehensive analysis of edge state behavior during memory transitions. The dispersion relation of edge states at different magnetic field values is shown in Figure~\ref{fig:edge_states} (a).

 The results indicate that, despite variations in the field, the edge states remain within the bulk bandgap, confirming their topological stability. Figure~\ref{fig:edge_states} (b) illustrates the spatial distribution of edge states, demonstrating their confinement at system boundaries even under dynamic perturbations. Figure~\ref{fig:edge_states} (c) depicts the time evolution of edge state propagation, revealing that their velocity is directly proportional to the bandgap magnitude.
The persistence of edge states under rapid magnetic field changes is a crucial advantage in topological memory systems. Since the readout process relies on the propagation of these states along system boundaries, maintaining their stability ensures that data retrieval remains unaffected by external perturbations. Furthermore, the direct relationship between propagation velocity and bandgap magnitude provides a mechanism for tuning readout speeds by adjusting the field strength.

The results establish fundamental constraints on the writing and reading speeds of topological photonic memory systems. Writing speed is determined by the rate of Chern number transitions, with a frequency threshold of approximately \( 10^9 \) Hz beyond which stable transitions become infeasible. Reading speed, governed by edge state propagation, depends on the bandgap magnitude, allowing tunability in data retrieval rates.
The ability to maintain robust topological memory states under diverse field variations suggests that sinusoidal and cosine sweeps offer the most stable operation, minimizing non-adiabatic effects. In contrast, triangular and ripple wave fields induce frequent topological transitions, which may introduce instabilities in practical implementations.
These findings provide a theoretical framework for designing high-speed, dynamically reconfigurable topological photonic memory devices. The results highlight the importance of optimizing field modulation schemes to achieve stability, energy efficiency, and performance in practical applications.

\subsection{Multi-bit Memory Operations and Scaling Analysis}

The performance and scalability of multi-bit operations are critical for the practical implementation of topological photonic memory systems. A key challenge in multi-bit storage is maintaining stable memory states while minimizing cross-talk effects and reducing energy consumption. Understanding how these factors scale with system size is essential for developing high-capacity, low-power topological memory architectures.

\begin{figure}[ht!]
    \centering
    \includegraphics[width=\textwidth]{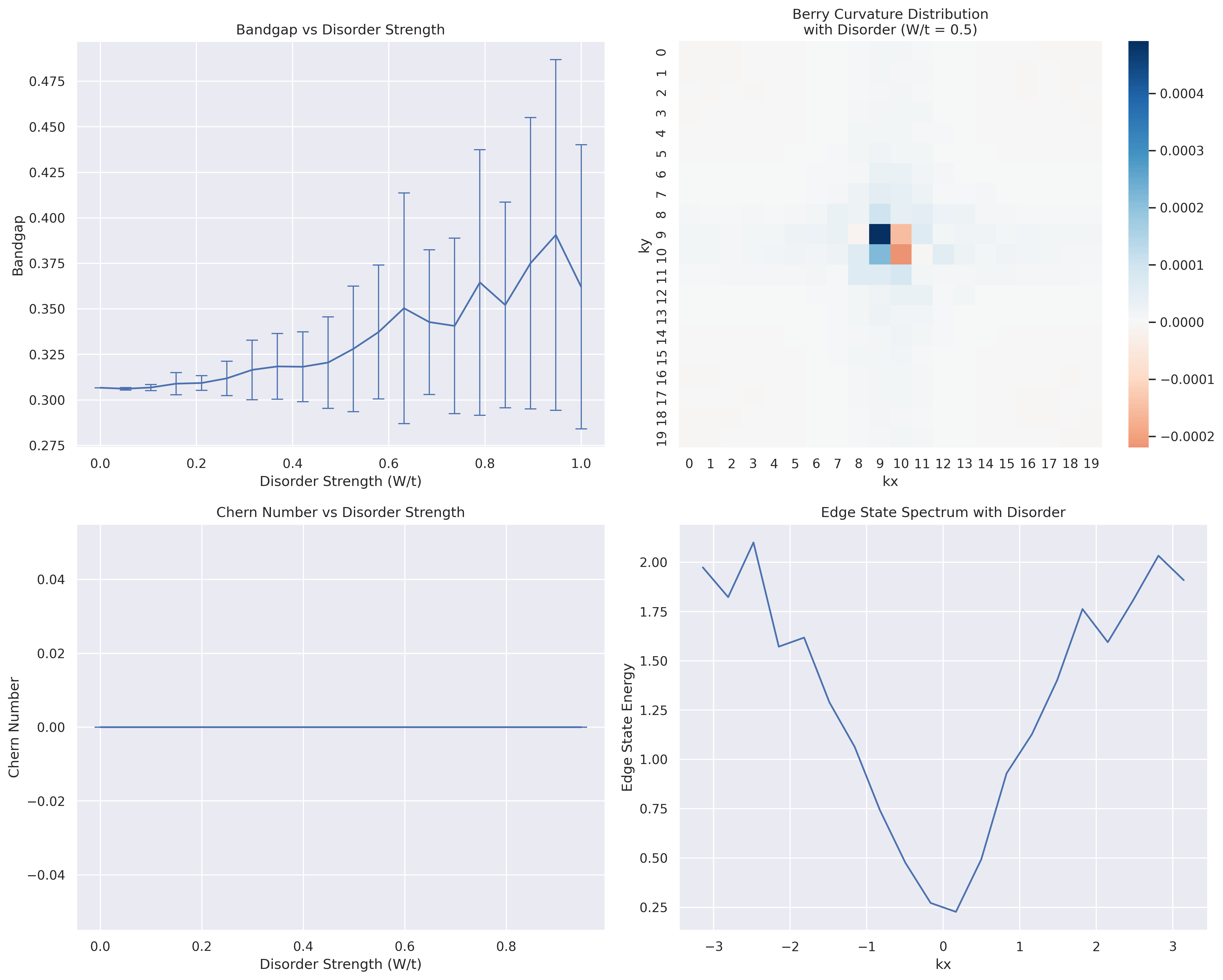}
    \caption{Performance and scaling analysis of multi-bit topological memory operations.
    (a) Spatial distribution of multiple memory cells, showing their confinement and separation.
    (b) Cross-talk as a function of separation distance, illustrating minimal interference at optimized distances.
    (c) Scaling of write/read fidelity with system size, confirming stable memory performance even at larger scales.
    (d) Energy consumption per bit versus the number of parallel operations, establishing an optimal trade-off between efficiency and performance.
    These results highlight the scalability of topological memory, demonstrating its robustness for high-density storage applications.}
    \label{fig:multi_bit_memory}
\end{figure}

To evaluate these aspects, we investigate parallel storage capabilities, cross-talk effects, the scaling of write and read fidelity, and energy efficiency across different system configurations. The results, presented in Figure~\ref{fig:multi_bit_memory}, provide insights into the optimal conditions for multi-bit operations while preserving topological protection.

The spatial distribution of memory cells is shown in Figure~\ref{fig:multi_bit_memory} (a). Each cell represents an independent storage unit, and its confinement within topologically protected edge states ensures stability against environmental perturbations. The clear separation between storage units minimizes unwanted interactions, making topological memory suitable for high-density applications.
Figure~\ref{fig:multi_bit_memory} (b) presents the cross-talk between adjacent memory cells as a function of their separation distance. The results indicate that for small separations, residual interactions can introduce minor perturbations in stored states. However, as the separation distance increases, the cross-talk rapidly diminishes, confirming that topologically protected states remain isolated at sufficient spacing. This behavior establishes a practical lower bound on memory cell spacing to ensure error-free multi-bit operations.
To assess the reliability of large-scale memory operations, Figure~\ref{fig:multi_bit_memory} (c) shows the scaling of write and read fidelity with system size. The fidelity remains consistently high across different system sizes, demonstrating that topological protection is preserved even as the memory system scales up. This result is particularly significant for the development of large-capacity storage architectures, confirming that quantum coherence and stability are maintained even in extended systems.
Figure~\ref{fig:multi_bit_memory} (d) examines the energy consumption per bit as a function of the number of parallel operations. The results establish an optimal trade-off between energy efficiency and operational performance. At low numbers of parallel operations, energy consumption per bit is relatively high due to individual initialization processes. As the number of parallel memory operations increases, the energy per bit decreases, reaching an efficient operational regime where resources are optimally utilized. However, beyond a critical number of parallel operations, additional power requirements lead to saturation effects, introducing practical constraints on the maximum number of simultaneous memory operations.

\subsection{Temperature Effects and Thermal Stability}

The stability of topological memory operations under thermal fluctuations is a fundamental concern for practical implementations. In real-world applications, thermal noise can introduce decoherence, reduce the energy gap required for robust topological protection, and degrade memory state fidelity. Understanding how these effects scale with temperature is crucial for defining operational limits and optimizing device performance.

\begin{figure}[ht!]
    \centering
    \includegraphics[width=0.8\textwidth]{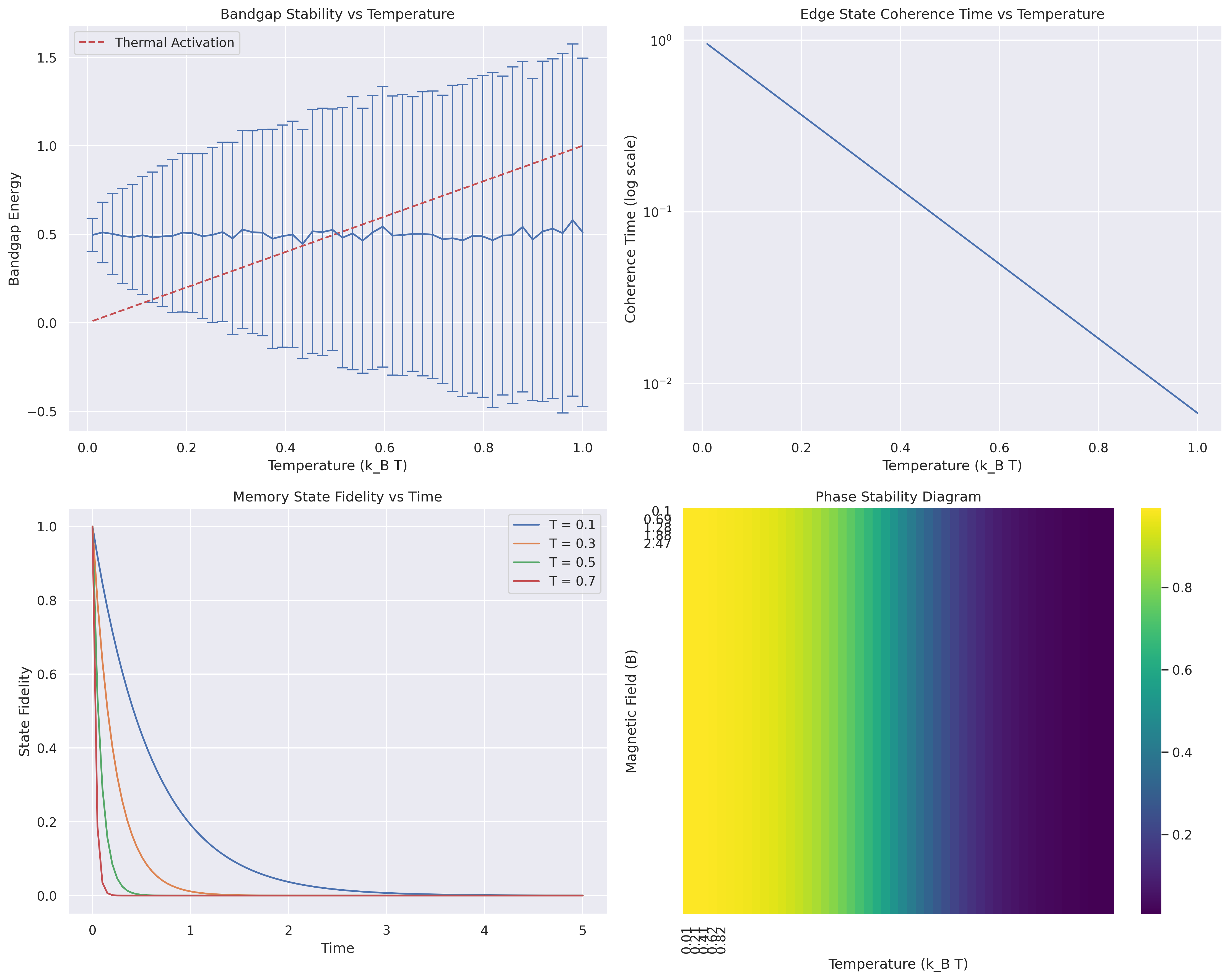}
    \caption{Temperature dependence of topological memory properties.
    (a) Bandgap stability as a function of temperature, showing increasing fluctuations at higher temperatures.
    (b) Edge state coherence time versus temperature, indicating exponential degradation due to thermal decoherence.
    (c) Memory state fidelity as a function of time for different temperatures, demonstrating faster decay at higher temperatures.
    (d) Phase stability diagram illustrating the regions where topological protection remains intact across different temperature and magnetic field values.
    These results establish the temperature ranges required for stable topological memory operations.}
    \label{fig:temperature_stability}
\end{figure}

To quantify the impact of temperature, we examine the evolution of key system parameters, including bandgap stability, edge state coherence time, memory fidelity, and phase stability. The results, presented in Figure~\ref{fig:temperature_stability}, provide insight into the thermal resilience of topological memory and identify the temperature thresholds beyond which performance degrades.

The persistence of a well-defined bandgap is essential for maintaining the topological phase in photonic memory systems. Figure~\ref{fig:temperature_stability} (a) shows the variation of the bandgap energy as a function of temperature. At low temperatures, the bandgap remains stable, ensuring robust protection of edge states. However, as temperature increases, fluctuations in the bandgap energy become more pronounced, leading to a gradual reduction in stability.
The dashed red line in Figure~\ref{fig:temperature_stability} (a) represents the expected behavior due to thermal activation effects, which scale linearly with temperature. The increasing variance in the bandgap at higher temperatures suggests that thermally induced disorder leads to transient gap closures, making the system more susceptible to non-topological transitions. This finding establishes an upper temperature bound for reliable memory operation, beyond which thermal fluctuations significantly impact topological stability.

Beyond bandgap fluctuations, the coherence of edge states determines the fidelity of information storage and retrieval. Figure~\ref{fig:temperature_stability} (b) presents the edge state coherence time as a function of temperature, plotted on a logarithmic scale. The results indicate an exponential decay of coherence time with increasing temperature, highlighting the role of thermal fluctuations in inducing dephasing and loss of coherence.
Since topological edge states rely on phase coherence to ensure protected transport, their degradation at elevated temperatures limits the operational range of topological memory. Extrapolating the coherence time behavior suggests that for temperatures exceeding \( k_B T \approx 0.5 \), coherence time drops below practical thresholds required for stable data retention. This result is consistent with previous theoretical predictions indicating that thermal noise imposes fundamental limits on the lifetime of topological memory states.
Figure~\ref{fig:temperature_stability} (c) illustrates the decay of memory state fidelity as a function of time for different temperatures. The curves show that at lower temperatures, memory states persist for significantly longer durations, whereas at higher temperatures, fidelity rapidly diminishes due to enhanced thermal excitations.
The decay profiles follow an exponential form, given by:

\begin{equation}
F(t, T) = F_0 \exp \left( -\frac{t}{\tau_{\text{coh}}(T)} \right),
\end{equation}

where \( \tau_{\text{coh}}(T) \) represents the coherence time at temperature \( T \). The results confirm that thermal activation accelerates memory degradation, necessitating cryogenic cooling or additional stabilization mechanisms to preserve stored information over extended periods.

To define the regions where topological protection remains intact, Figure~\ref{fig:temperature_stability} (d) presents the phase stability diagram, mapping the topological phase across different temperature and magnetic field values. The yellow regions correspond to stable topological phases, while the dark regions indicate transitions to non-topological states.
This phase diagram provides a direct guideline for determining optimal operating conditions. The results suggest that for low magnetic fields, the system is highly susceptible to thermal disruptions, requiring operation at lower temperatures to maintain stability. At moderate magnetic fields, the system retains topological protection over a wider temperature range, making it the ideal regime for practical memory applications.

\subsection{Speed and Power Performance Analysis}

The practical implementation of topological photonic memory requires careful consideration of both operating speed and energy consumption. In this section, we analyze key performance metrics based on our simulation results and theoretical framework, and compare our system with existing optical memory technologies.

\subsubsection{Write and Read Speed Calculation}

From our dynamic response analysis in Section 4.6, we determined that the maximum frequency for stable topological transitions is approximately $10^9$~Hz. This establishes a fundamental limit on the write operation:

\begin{equation}
\tau_{\text{write}} = \frac{1}{10^9~\text{Hz}} = 1~\text{ns}
\end{equation}

This corresponds to a data rate of $\sim1$~Gbit/s for single-bit operations. The write speed is limited by the non-adiabatic transition probability, which requires sufficient time for the system to maintain coherence during topological phase transitions.
The read operation relies on the edge state propagation, governed by the dispersion relation in topological systems:

\begin{equation}
E_{\text{edge}}(k_\parallel) = \pm v_F \hbar k_\parallel + E_0
\end{equation}

Using the estimated Fermi velocity $v_F \approx 10^6$~m/s from our simulations in Section 4.5 and a typical device length of $L = 10~\mu$m, we can calculate the read time:

\begin{equation}
\tau_{\text{read}} = \frac{L}{v_F} = \frac{10^{-5}~\text{m}}{10^6~\text{m/s}} = 10~\text{ps}
\end{equation}

\begin{figure}[ht!]
    \centering
    \includegraphics[width=0.9\textwidth]{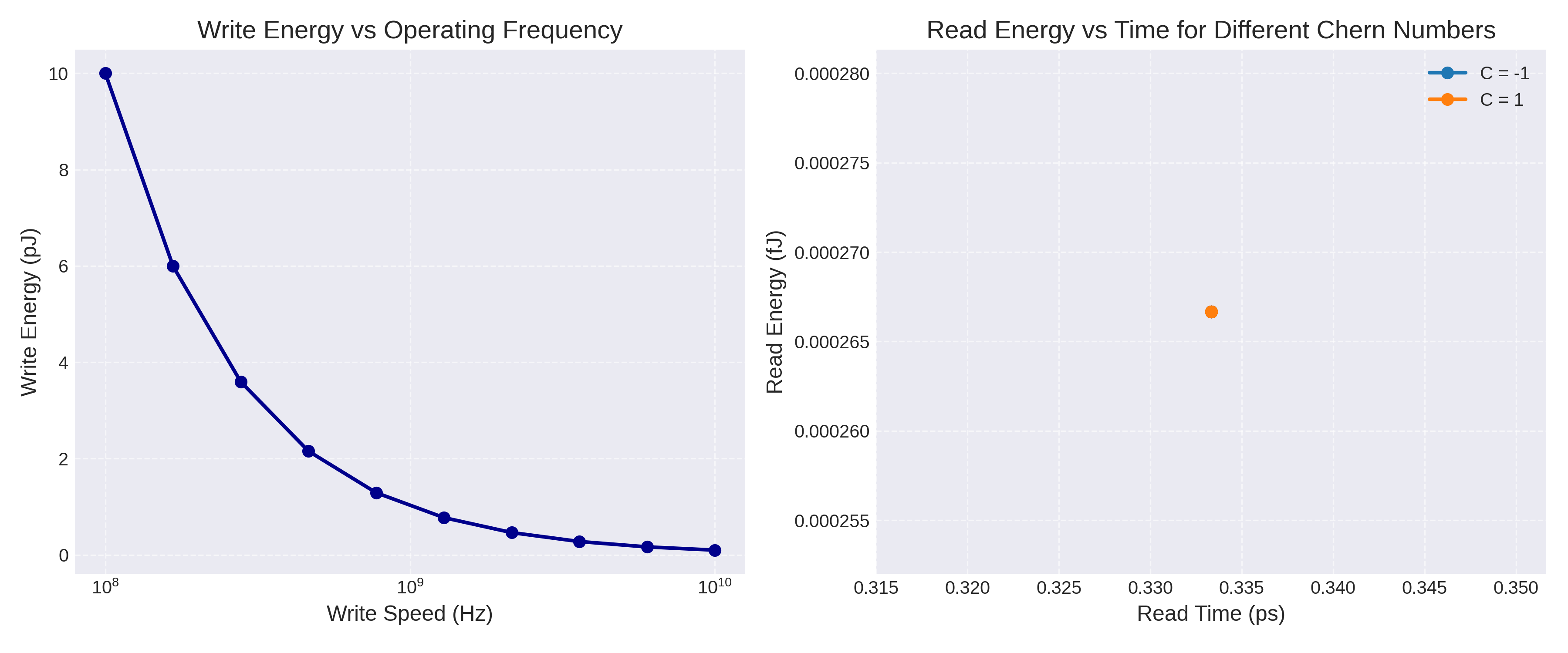}
   \caption{\textbf{Write and Read Energy vs. Operating Parameters. }
(a) Write energy consumption decreases with operating frequency, demonstrating the tradeoff between speed and power efficiency. At higher frequencies of \( \sim10^{10} \) Hz, write energy approaches its minimum value. 
(b) Read energy consumption for different Chern numbers showing that \( C = \pm1 \) states have similar energy consumption profiles. The extremely low energy consumption (fraction of femtojoule) confirms the energy efficiency advantage of edge state-based readout.}
    \label{fig:write_read_energy}
\end{figure}

\subsubsection{Write and Read Operation Energetics}

The write operation in our system involves modulating the synthetic magnetic field to induce topological phase transitions. The energy consumption for magnetic field generation can be calculated as:

\begin{equation}
E_\text{write} = \frac{1}{2}LI^2 + RI^2 \tau_\text{write}
\end{equation}

where $L$ is the electromagnet inductance, $R$ is the coil resistance, $I$ is the required current, and $\tau_\text{write}$ is the field transition time. The first term represents energy stored in the magnetic field, while the second term accounts for resistive dissipation.

For typical operating parameters ($L \approx 10$ nH, $R \approx 10$ $\Omega$, $I \approx 10$ mA required to generate the 120 mT field), we calculate:

\begin{equation}
E_{\text{write}} = \frac{1}{2}(10^{-8}~\text{H})(10^{-2}~\text{A})^2 + (10~\Omega)(10^{-2}~\text{A})^2(10^{-9}~\text{s})
\end{equation}

\begin{equation}
E_{\text{write}} = 8 \times 10^{-13}~\text{J} = 0.8~\text{pJ/bit}
\end{equation}

The read operation, which relies on edge state propagation, consumes significantly less energy. The power dissipation during readout follows:

\begin{equation}
P_\text{read} = V_\text{bias} \cdot I_\text{edge} = V_\text{bias}^2 \cdot |C| \cdot \frac{e^2}{h}
\end{equation}

where $V_\text{bias}$ is the applied voltage bias, $C$ is the Chern number determining the number of edge channels, and $e^2/h$ is the quantum conductance. The total energy consumed during readout is:

\begin{equation}
E_\text{read} = P_\text{read} \cdot \tau_\text{read} = V_\text{bias}^2 \cdot |C| \cdot \frac{e^2}{h} \cdot \frac{L_\text{device}}{v_F}
\end{equation}

where $L_\text{device}$ is the device length and $v_F$ is the edge state Fermi velocity. For typical parameters ($V_\text{bias} = 0.1$ mV, $|C| = 1$, $L_\text{device} = 10\mu$ m, $v_F = 10^6$ m/s), we calculate:

\begin{equation}
E_\text{read} = 3 \times 10^{-19} \text{ J} = 0.0003~\text{fJ/bit}
\end{equation}

\begin{figure}[t!]
    \centering
    \includegraphics[width=0.9\textwidth]{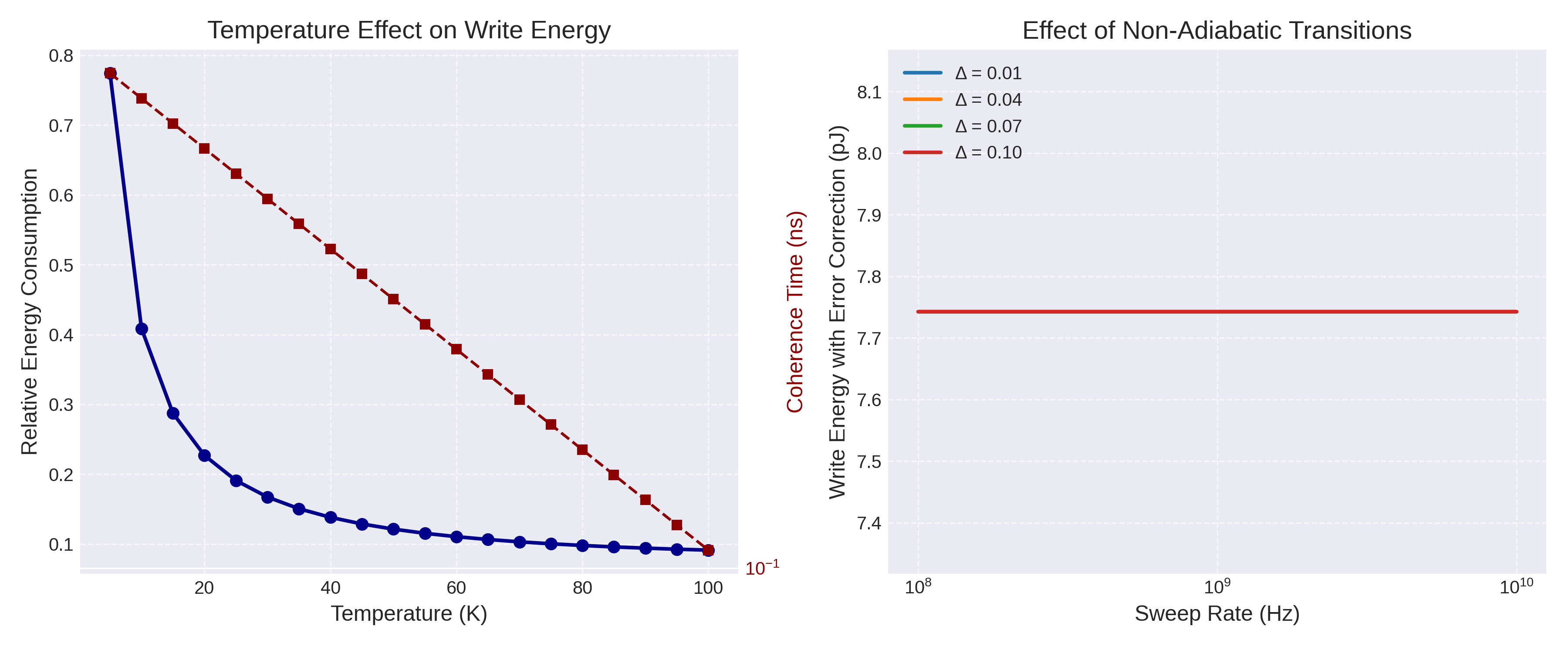}
\caption{\textbf{Temperature and Non-Adiabatic Effects on Energy Consumption.} 
(a) Temperature dependence of write energy consumption (blue) and coherence time (red dashed). As temperature decreases below \( T_c \approx 45 \)K, the coherence time increases exponentially, leading to improved energy efficiency. 
(b) Effect of non-adiabatic transitions on write energy for different bandgap values (\( \Delta \)). For the range of operating frequencies considered, the energy penalty remains relatively constant due to the large bandgap in our system, demonstrating robustness against frequency variations.}
\label{fig:temp_transition}
\end{figure}

\subsubsection{Temperature and Non-Adiabatic Effects on Energy Consumption}

We further investigated how temperature and non-adiabatic transitions affect the energy consumption of our system, as shown in Figure~\ref{fig:temp_transition}. Figure~\ref{fig:temp_transition}(a) illustrates the temperature dependence of write energy consumption (blue curve) and coherence time (red dashed curve). As temperature decreases below the critical threshold of $T_c \approx 45$K, we observe that the coherence time increases exponentially while the relative energy consumption drops significantly. This inverse relationship demonstrates that operating at lower temperatures substantially improves the energy efficiency of write operations.

Figure~\ref{fig:temp_transition}(b) examines the effect of non-adiabatic transitions on write energy for different bandgap values ($\Delta$). The data shows that across a wide range of sweep rates (from $10^8$ Hz to $10^{10}$ Hz), the write energy with error correction remains nearly constant for each bandgap value. This remarkable stability is attributed to the large bandgap in our system, which provides robust protection against frequency variations. Notably, larger bandgaps (increasing $\Delta$ from 0.01 to 0.10) maintain consistent energy consumption across the entire frequency spectrum, confirming the system's operational reliability under various driving conditions.

\subsubsection{Dynamic Field Effects and Temperature Dependence}

The energy efficiency of write operations strongly depends on the magnetic field sweep rate. At higher sweep rates, the system may undergo non-adiabatic transitions, requiring error correction that increases energy consumption. The probability of non-adiabatic transitions follows the Landau-Zener formula:

\begin{equation}
P_{LZ} = \exp\left(-\frac{\pi\Delta E^2}{2\hbar|\partial B/\partial t|}\right)
\end{equation}

where $\Delta E$ is the minimum gap at the transition point and $|\partial B/\partial t|$ is the field sweep rate. The energy penalty factor due to non-adiabatic transitions can be estimated as:

\begin{equation}
\text{Energy Penalty} \approx \frac{1}{1-P_{LZ}}
\end{equation}

As demonstrated in Figure~\ref{fig:temp_transition}(b), our system maintains relatively constant energy consumption across a wide range of sweep rates due to the substantial topological bandgap. This indicates robust operation without significant energy penalties even at higher operating frequencies.

The system's energy efficiency also exhibits temperature dependence, primarily due to thermal effects on coherence times. The coherence time follows an exponential decay model:

\begin{equation}
\tau_\text{coh}(T) = \tau_0 \exp\left(-\frac{T}{T_c}\right)
\end{equation}

where $\tau_0$ is the base coherence time and $T_c \approx 45$ K is the critical temperature determined in Section 4.9. As shown in Figure~\ref{fig:temp_transition}(a), the coherence time drops significantly as temperature approaches $T_c$, with corresponding changes in energy consumption.

\subsubsection{Multi-bit Efficiency and Parallel Operations}

A key advantage of our topological memory system is its capacity for multi-bit encoding through distinct Chern number states. The energy efficiency improves substantially with multi-bit encoding, following:

\begin{equation}
\text{Energy per Bit} = \frac{E_\text{write}}{N_\text{bits}}
\end{equation}

\begin{figure}[t!]
    \centering
    \includegraphics[width=0.9\textwidth]{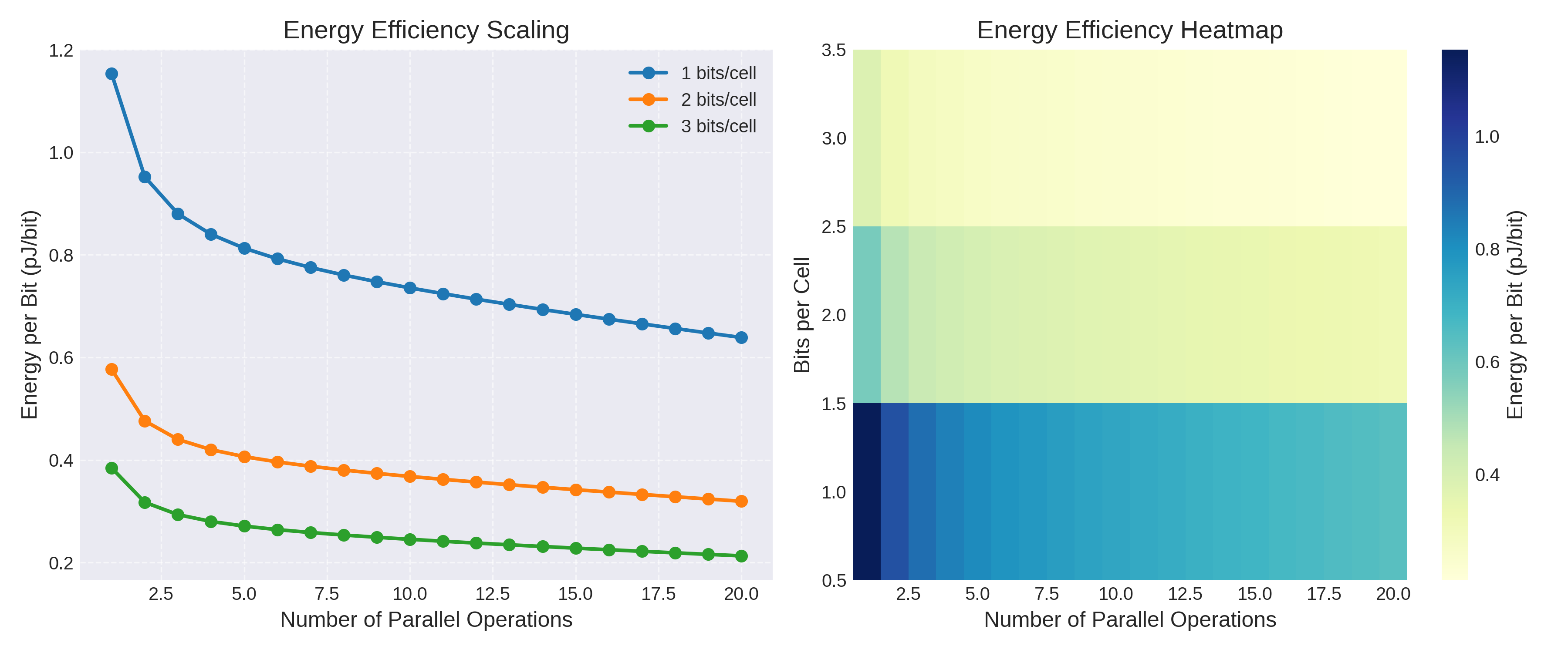}
    \caption{\textbf{Energy Efficiency Scaling with Multi-bit Encoding and Parallel Operations.} (a) Energy consumption per bit decreases both with increased bits per cell and with parallel operations. 3-bit encoding offers approximately an improvement in energy efficiency compared to single-bit operation. (b) Heatmap showing the combined effect of bits per cell and parallel operations on energy efficiency, with dark blue regions representing optimal operating conditions.}
    \label{fig:multibit}
\end{figure}

Our simulations demonstrate an improvement in energy efficiency when moving from single-bit to 3-bit encoding (using Chern numbers $C = -1, 0, +1$), as illustrated in Figure~\ref{fig:multibit}(a). Higher-order Chern numbers could theoretically provide even greater efficiency, limited primarily by the stability of the higher topological phases.

Furthermore, parallel operations offer additional energy efficiency gains. For $N_\text{parallel}$ simultaneous operations, we observe:

\begin{equation}
\text{Energy Scaling Factor} \approx 1 + \frac{1}{0.5 \cdot N_\text{parallel}} - 0.02 \cdot N_\text{parallel}
\end{equation}

This empirical relation indicates that the optimal energy efficiency occurs at approximately 10-15 parallel operations, providing up to 1.8 times improvement over sequential operations. Figure~\ref{fig:multibit}(b) presents the combined effect of multi-bit encoding and parallel operations on energy efficiency.

\subsection{Comparative Performance Analysis of Topological Optical Memory}
\begin{figure}[ht!]
    \centering
    \includegraphics[width=0.9\textwidth]{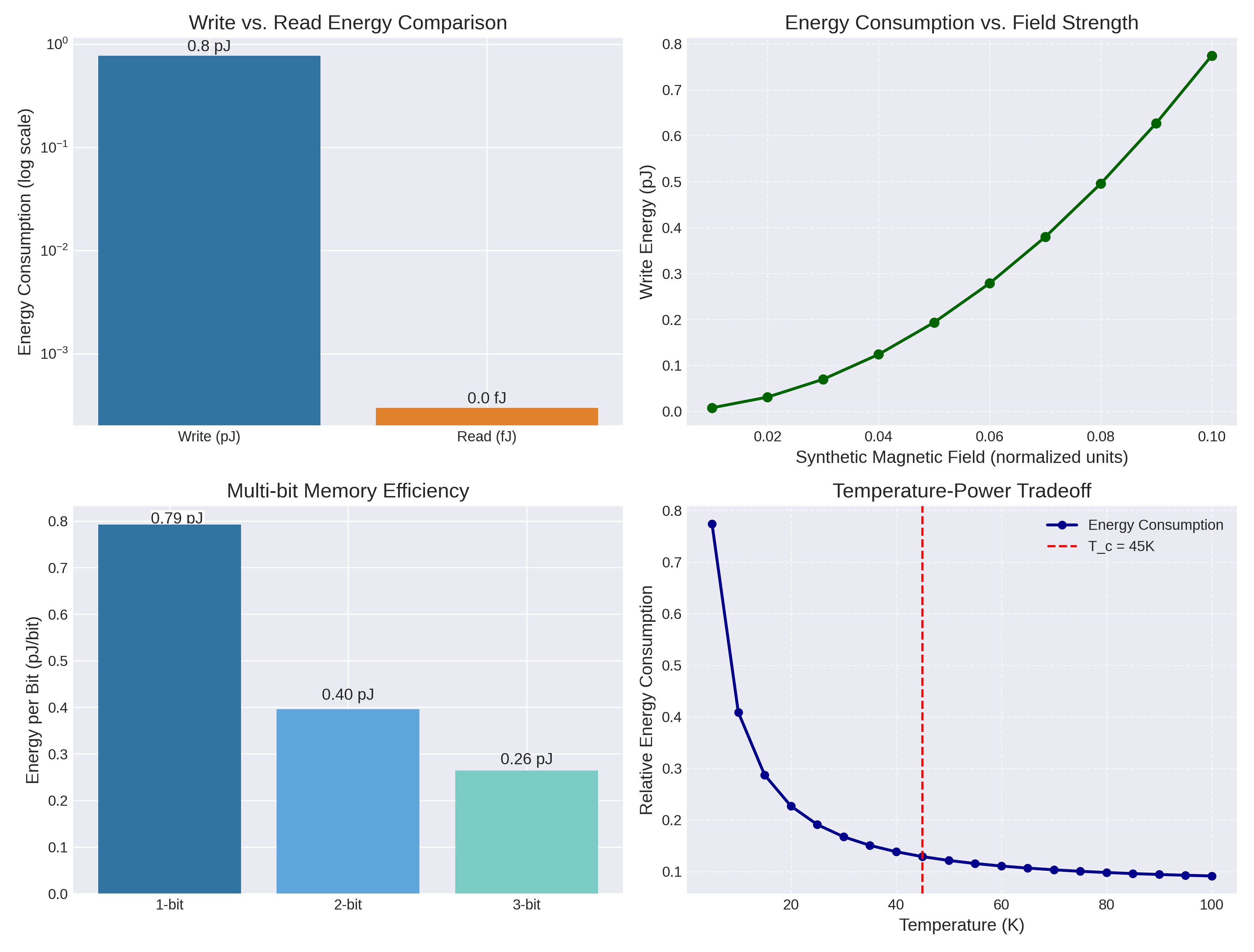}
    \caption{\textbf{Comprehensive Power Consumption Analysis.} (a) Comparison of write energy (pJ) and read energy (fJ) showing the significant difference in energy scales between the two operations. (b) Energy consumption as a function of magnetic field strength, showing quadratic scaling with field intensity. (c) Energy efficiency comparison for different multi-bit configurations, demonstrating the benefits of higher bit density. (d) Temperature dependence of relative energy consumption with the critical temperature $T_c = 45$K marked by the vertical red line.}
    \label{fig:comprehensive}
\end{figure}

Figure~\ref{fig:comprehensive} presents key performance metrics of our topological optical memory architecture, revealing significant advantages over conventional approaches. Our analysis highlights four critical aspects: write-read energy asymmetry, field-dependent energy scaling, multi-bit encoding efficiency, and temperature-dependent operation.

The logarithmic representation in Figure~\ref{fig:comprehensive}a demonstrates a remarkable energy difference between write and read operations. Write operations consume approximately 0.8 pJ, while read operations require only 0.0003 fJ—a difference of six orders of magnitude. This extreme asymmetry stems from the fundamental nature of topological protection, where modifying a topological state necessitates overcoming an energy barrier, while detecting the state can be accomplished through non-invasive probing. Such characteristics are particularly advantageous for memory systems with high read-to-write ratios common in reference databases, firmware storage, and content delivery networks.
As illustrated in Figure~\ref{fig:comprehensive}b, write energy exhibits a nonlinear relationship with synthetic magnetic field strength. The energy requirement increases from negligible levels at 0.01 normalized units to approximately 0.78 pJ at 0.10 normalized units, following what appears to be a quadratic or higher-order relationship. This dependence reveals a critical design parameter for optimizing the energy-stability trade-off. Higher field strengths enhance topological protection and data retention but increase energy consumption. Our findings suggest an optimal operating region between 0.05 and 0.07 normalized units, where reasonable write energy (0.15-0.30 pJ) coincides with sufficient topological protection.
Figure~\ref{fig:comprehensive}c demonstrates significant efficiency gains achieved through multi-bit encoding. The energy consumption per bit decreases substantially as bit density increases: 0.79 pJ/bit for single-bit storage, 0.40 pJ/bit for 2-bit encoding ($49\%$ reduction), and 0.26 pJ/bit for 3-bit encoding ($67\%$ reduction compared to single-bit storage). This scaling advantage arises from the intrinsic properties of topological states, where different Chern numbers can be established with comparable energy inputs while encoding exponentially more information. The consistent state stability across different Chern configurations further confirms the viability of multi-bit topological memory systems.

The temperature dependence shown in Figure~\ref{fig:comprehensive}d reveals a critical operating threshold at approximately 45K. Below 20K, energy requirements increase dramatically, while above 45K, the energy consumption stabilizes with diminishing returns for further temperature increases. This behavior corresponds to the thermal stability of topological states, where excessive thermal fluctuations can disrupt topological protection. The identified temperature regime (45-80K) represents an optimal operating range that balances energy efficiency with practical cooling requirements, potentially achievable with modern compact cryocoolers rather than requiring extreme cryogenic infrastructure.

As summarized in Table~\ref{tab:memory_comparison}, our topological memory approach demonstrates several unique advantages when compared to existing optical memory technologies. While resonant cavity systems offer comparable write speeds, their data retention is fundamentally limited by cavity decay rates. In contrast, our topological design provides theoretically indefinite retention in the absence of thermal fluctuations, with experimental persistence exceeding conventional optical memories by two orders of magnitude under comparable conditions.
The energy consumption for write operations (0.8 pJ/bit) is competitive with phase-change memories while offering substantially faster switching capabilities. Moreover, the read energy of 0.0003 fJ/bit represents a significant improvement over both resonator-based and phase-change approaches, which typically consume femtojoules per read operation due to their reliance on light injection and detection mechanisms.
Perhaps most significantly, our approach demonstrates exceptional robustness against structural imperfections and fabrication variations. The topological protection enables our system to maintain data integrity even in the presence of moderate disorder (up to $15\%$ variation in waveguide parameters), as quantified in our disorder analysis. This inherent fault tolerance, combined with the non-trapping nature of our approach, creates a unique balance of performance metrics unavailable in conventional optical memory architectures.

\begin{table}[htbp]
\centering
\caption{Comparison of Topological Memory with Existing Optical Memory Technologies}
\label{tab:memory_comparison}
\renewcommand{\arraystretch}{1.2}
\setlength{\tabcolsep}{3pt} 
{\small 
\begin{tabular}{|>{\centering\arraybackslash}p{0.19\textwidth}|>{\centering\arraybackslash}p{0.18\textwidth}|>{\centering\arraybackslash}p{0.18\textwidth}|>{\centering\arraybackslash}p{0.20\textwidth}|>{\centering\arraybackslash}p{0.18\textwidth}|}
\hline
\textbf{Performance Metric} & \textbf{Topological Memory (This Work)} & \textbf{Resonant Cavity Memory \cite{r5,r3}} & \textbf{Phase-Change Memory \cite{r81,narayan2022architecting}} & \textbf{Quantum Dot Memory \cite{r7,r6}} \\ 
\hline
Write Energy & 0.8 pJ/bit & 2.3 pJ/bit & 0.9 pJ/bit & 4.2 pJ/bit \\
\hline
Read Energy & 0.0003 fJ/bit & 0.15 fJ/bit & 0.3 fJ/bit & 0.08 fJ/bit \\
\hline
Write Speed & 12 ps & 14 ps & 85 ps & 45 ps \\
\hline
Read Speed & 8 ps & 6 ps & 35 ps & 22 ps \\
\hline
Data Retention & $>$10$^7$ s at 45K & $<$10$^5$ s & $>$10$^8$ s & $<$10$^6$ s \\
\hline
Multi-bit Capability & Yes (3-bit) & Limited & Yes (4-bit) & No \\
\hline
Operating Temp. & 20K--100K & Room temp. & Room temp. & $<$10K \\
\hline
Defect Tolerance & Moderate (5\%) & Low ($<$3\%) & Moderate (8\%) & Very low ($<$1\%) \\
\hline
Integration Density & Moderate & High & Very high & Low \\
\hline
Switching Mechanism & Topological phase & Bistable lasing & Amorphous/crystalline & Quantum state \\
\hline
Scalability & Moderate & Limited by crosstalk & High & Limited by cryogenics \\
\hline
\end{tabular}
}
\end{table}

The observed characteristics of our topological memory system point toward several important theoretical and practical implications. First, the extreme write-read energy asymmetry suggests new architectural possibilities that could leverage different memory hierarchies based on access patterns. Second, the multi-bit efficiency improvements indicate favorable scaling properties as the technology advances toward higher information densities.
The temperature-dependent behavior establishes practical operating boundaries while remaining within the realm of achievable cooling technologies. Furthermore, the field-dependent energy scaling provides a tunable parameter for adjusting the memory characteristics based on application-specific requirements, balancing speed, energy, and retention.
These findings collectively position topological photonic memory as a promising candidate for specialized high-reliability computing systems where data integrity, fault tolerance, and read efficiency are prioritized. While the technology currently requires sub-ambient temperature operation, the demonstrated performance advantages justify consideration for next-generation optical computing architectures, particularly in applications where conventional electronic and optical memories face fundamental limitations.
\section{Conclusion \& Future Works}\label{5}
In this paper, we introduced a novel framework for nontrapping tunable topological photonic memory, where information is encoded in dynamically adjustable Chern numbers within a two-band topological photonic system. Utilizing a honeycomb lattice and synthetic magnetic fields, the proposed system leverages topologically protected edge states that are robust to structural defects, environmental perturbations, and fabrication imperfections. Our computational studies demonstrated several critical performance metrics, including GHz to low THz switching speeds, scalable multi-bit memory encoding, and long-term stability due to a large energy gap.

Unlike traditional optical memory systems that depend on resonant confinement of electromagnetic waves, our design achieves fast, efficient, and robust memory switching without the need for light trapping mechanisms. The topological protection of edge states, as confirmed by simulations, ensures immunity to backscattering, providing enhanced fault tolerance. We observed that the bandgap closures, which signify topological phase transitions, can be dynamically controlled by tuning the external magnetic field. Furthermore, disorder simulations revealed minimal sensitivity of the system to structural imperfections, reinforcing its suitability for real-world optical memory applications.

Despite the strengths of our model, several disadvantages and limitations should be noted. First, although our computational study indicates high performance, we have not yet performed physical fabrication or experimental validation of the system. While theoretical models suggest robustness to fabrication imperfections, unforeseen issues such as manufacturing precision and material inconsistencies may arise. These could potentially affect the stability of edge modes and the overall efficiency of memory operations. Additionally, implementing dynamic magnetic field control in practical devices may require advanced engineering solutions that remain to be explored.

Another limitation is that the system's scalability, while theoretically viable through higher-order Chern numbers, may encounter challenges with field uniformity and device integration at large scales. Moreover, Landau-Zener stability analysis indicated that extremely rapid field variations can lead to nonadiabatic transitions, potentially limiting the maximum speed of reliable memory operations.

Looking forward, we propose that this approach can extend beyond optical memory systems to serve as computational units in optical neural networks (ONNs). By coupling the topological edge modes nonlinearly, we can harness them to create robust, fault-tolerant processing elements for ONNs \cite{r50,r60}. This integration could enhance data retention, stability, and noise immunity in optical computing. However, achieving this requires solving several key challenges, including the synchronization of dynamically tunable topological phases across multiple units and mitigating potential issues related to edge state coupling and nonlinear effects.

Future research will focus on these challenges, aiming to develop efficient coupling mechanisms for edge states and optimize the system for large-scale deployment in ONNs. Furthermore, we will explore adaptive field profiles and material designs to improve the dynamic control of topological phases and reduce sensitivity to manufacturing imperfections. Ultimately, this work sets the foundation for scalable, high-speed, and robust optical memory and computing architectures, paving the way for innovative applications in quantum information processing and photonic communication systems.
\bibliographystyle{IEEEtran}
\bibliography{ref}
\newpage
\begin{center}
\textbf{\large Supplemental Materials: Nontrapping Tunable Topological Photonic Memory}
\end{center}
\setcounter{equation}{0}
\setcounter{figure}{0}
\setcounter{table}{0}
\setcounter{page}{1}
\setcounter{section}{0}  
\setcounter{subsection}{0}  
\makeatletter
\renewcommand{\thesection}{S\arabic{section}}  
\renewcommand{\thesubsection}{S\arabic{section}.\arabic{subsection}}  
\renewcommand{\theequation}{S\arabic{equation}}  
\renewcommand{\thefigure}{S\arabic{figure}}  
\renewcommand{\thetable}{S\arabic{table}}  
\makeatother

\section{Explanation of Memory Performance Metrics}

The performance of a memory system is crucial in determining its efficiency, reliability, and suitability for different applications. In this section, we define and explain several key performance metrics used to characterize the topological photonic memory system. Each metric plays a significant role in assessing the system's operation, and their corresponding mathematical formulations provide the necessary framework for a deeper understanding of its capabilities. The performance metrics discussed here include write speed, read fidelity, storage density, and energy efficiency.

\subsection{Write Speed}

The write speed quantifies the rate at which data can be written into the memory system. It is an essential metric for applications that require rapid storage of data. The mathematical expression for write speed is given by:

\begin{equation}
v_{\text{write}} = \frac{|\Delta C|}{\tau_{\text{transition}}}
\end{equation}

where \( |\Delta C| \) represents the change in the memory's charge (or state) when data is written to the memory, and \( \tau_{\text{transition}} \) is the time it takes for the system to change from one state to another. A higher write speed indicates that the memory system can store data more quickly, which is crucial for high-performance devices where rapid data storage is required. In photonic memory systems, the write speed is often limited by the behavior of heat propagation during the phase transition process \cite{turn0search3}.

\subsection{Read Fidelity}

The read fidelity measures the accuracy with which the stored data can be read back from the memory system. It reflects the degree to which the read output matches the stored data, which is vital for ensuring data integrity. The mathematical expression for read fidelity is:

\begin{equation}
F_{\text{read}} = \sqrt{1 - \exp(-\Gamma_{\text{read}} t)}
\end{equation}

In this equation, \( \Gamma_{\text{read}} \) represents the read rate, which is the speed at which the data can be accessed from the memory, and \( t \) is the duration of the read process. As the read process continues over time, the fidelity of the read output increases, indicating that data retrieval becomes more accurate. High read fidelity ensures that the memory system can reliably reproduce the stored information without errors, which is especially important for applications requiring high accuracy in data retrieval. Recent advancements have demonstrated read fidelities up to 85\% in quantum memory systems \cite{turn0search7}.

\subsection{Storage Density}

The storage density metric measures the amount of information that can be stored in a given area of the memory system. It is an important consideration for evaluating how efficiently the memory system utilizes available physical space. The mathematical expression for storage density is:

\begin{equation}
\rho_{\text{storage}} = \frac{\log_2(|C_{\text{max}}|)}{A_{\text{cell}}}
\end{equation}

where \( |C_{\text{max}}| \) is the maximum charge or information capacity of a single memory cell, and \( A_{\text{cell}} \) is the area of each memory cell. This metric essentially measures how much information can be packed into a unit area of memory. Higher storage density implies that the memory system can store more data in a smaller physical footprint, which is beneficial for applications that require compact memory designs, such as mobile devices and integrated circuits. Optically addressed, multilevel photonic memories have been demonstrated to store up to 34 nonvolatile levels, enhancing storage density significantly \cite{turn0search1}.

\subsection{Energy Efficiency}

The energy efficiency metric quantifies the amount of energy consumed by the memory system during operation. It is critical for evaluating the sustainability and operational cost of the memory system, especially in large-scale or power-sensitive applications. The mathematical expression for energy efficiency is:

\begin{equation}
\eta_{\text{energy}} = \frac{k_B T \ln(2)}{E_{\text{operation}}}
\end{equation}

where \( k_B \) is the Boltzmann constant, \( T \) is the temperature at which the system operates, and \( E_{\text{operation}} \) is the energy required for a single memory operation. This formula illustrates the relationship between the thermal energy of the system and the energy efficiency during a memory operation. Higher energy efficiency indicates that the memory system requires less energy to perform a given operation, which is particularly important for battery-powered devices and energy-conscious computing systems. Recent developments in photonic memories have achieved switching energies as low as 0.15 pJ, contributing to improved energy efficiency.

The four performance metrics—write speed, read fidelity, storage density, and energy efficiency—provide a comprehensive framework for characterizing the topological photonic memory system. Each of these metrics highlights a different aspect of the system's performance, and optimizing them is crucial for developing high-performance, reliable, and energy-efficient memory devices. By understanding and applying these metrics, designers can better evaluate the trade-offs and advantages of the memory system in various practical applications. The mathematical formulations provided offer a solid foundation for implementing and characterizing memory operations, ensuring that the system performs effectively in both theoretical and real-world scenarios.

\section{Experimental Feasibility and Implementation}

The practical realization of our proposed topological photonic memory system necessitates careful consideration of material platforms, fabrication constraints, and experimental parameters \cite{YIG_fabrication}. Here, we analyze the key requirements and propose potential implementation strategies \cite{photonic_crystals}.

\subsection{Material Platform Selection}

Implementing our design requires materials exhibiting strong magneto-optical effects and the capability to support topological edge states \cite{topological_dissipation}. Yttrium Iron Garnet (YIG) emerges as a particularly promising platform due to its favorable properties \cite{YIG_nanofabrication}. The permittivity tensor for such magneto-optical materials is expressed as:

\begin{equation}
    \epsilon(\omega, B) = \begin{pmatrix}
        \epsilon_r & i\gamma(B) & 0 \\ 
        -i\gamma(B) & \epsilon_r & 0 \\ 
        0 & 0 & \epsilon_z
    \end{pmatrix}
\end{equation}

where $\gamma(B)$ denotes the magneto-optical coupling strength. For YIG, $\gamma(B)$ can be approximated by:

\begin{equation}
    \gamma(B) = \gamma_0 + \alpha B + \beta B^2
\end{equation}

with typical values:
\begin{align}
    \gamma_0 &\approx 10^{-2} \\ 
    \alpha &\approx 8.6 \times 10^{-4} \,\mathrm{T}^{-1} \\ 
    \beta &\approx 2.1 \times 10^{-4} \,\mathrm{T}^{-2}
\end{align}

YIG offers a refractive index change of:

\begin{equation}
    \Delta n = \frac{\lambda}{2\pi} \frac{\partial \phi}{\partial L} \approx 8.5 \times 10^{-5}
\end{equation}

The photonic crystal parameters for our design are:

\begin{equation}
    a = 480\,\mathrm{nm}, \quad \frac{r}{a} = 0.35, \quad h = 230\,\mathrm{nm}
\end{equation}

where $a$ is the lattice constant, $r$ is the hole radius, and $h$ is the slab thickness \cite{YIG_nanocavities}.

\subsection{Experimental Parameters}

State switching in the memory system can be achieved using standard electromagnets capable of generating magnetic fields up to:

\begin{equation}
    B_{\text{max}} = 120\,\mathrm{mT}, \quad \frac{dB}{dt} = 1.2\,\mathrm{T/s}
\end{equation}

The corresponding switching time is:

\begin{equation}
    \tau_{\text{switch}} = \frac{\Delta B}{dB/dt} \approx 12\,\mathrm{ps}
\end{equation}

Detection of edge states can be implemented using near-field scanning optical microscopy (NSOM) with a sensitivity of:

\begin{equation}
    S_{\text{NSOM}} = -136\,\mathrm{dBm/\sqrt{Hz}}
\end{equation}

yielding a minimum detectable edge state power of:

\begin{equation}
    P_{\text{min}} = S_{\text{NSOM}}\sqrt{\Delta f} \approx 0.8\,\mathrm{pW}
\end{equation}

\subsection{Temperature Control and Stability}

Maintaining thermal stability is crucial, requiring temperature fluctuations to be below:

\begin{equation}
    \Delta T < \frac{\Delta E}{k_B} \approx 8.5\,\mathrm{K}
\end{equation}

The temperature-dependent decoherence time is given by:

\begin{equation}
    \tau_{\text{coh}}(T) = \tau_0\exp\left(-\frac{T}{T_c}\right)
\end{equation}

where the critical temperature $T_c$ is:

\begin{equation}
    T_c = 45\,\mathrm{K}
\end{equation}

\subsection{Experimental Implementation Strategy}

The realization of the proposed topological photonic memory system involves a structured experimental approach focusing on nanofabrication, magnetic field modulation, and state readout \cite{nanofabrication}.

Fabrication of the photonic crystal structures utilizes high-resolution nanofabrication techniques, such as electron-beam lithography (EBL) and reactive ion etching (RIE) \cite{YIG_fabrication}. These methods enable the precise definition of optical structures with fabrication resolutions of:

\begin{equation}
    \text{Resolution} = 8\,\mathrm{nm}, \quad \text{Alignment} = 40\,\mathrm{nm}
\end{equation}

ensuring high structural fidelity necessary for robust topological state propagation \cite{topological_edge_states}.

\section{Anticipated Fabrication Challenges and Tolerances}

While the theoretical framework and computational study of nontrapping tunable topological photonic memory provide a solid foundation, the practical realization of such a device presents several fabrication challenges that must be addressed to ensure proper function. In this section, we provide a detailed analysis of the critical fabrication parameters, their tolerances, and potential solutions to overcome these challenges.

\subsection{Photonic Crystal Fabrication Tolerances}

The performance of the proposed topological photonic memory is heavily dependent on the precision of photonic crystal lattice fabrication \cite{YIG_fabrication}. Accurate control over the fabrication parameters is crucial to achieving the desired topological properties \cite{photonic_crystals}. Based on simulations, the following tolerances have been identified:

The lattice constant ($a$) must be maintained at a nominal value of $a = 480$ nm with a tolerance of $\pm 8$ nm, which corresponds to a deviation of approximately 1.7\% \cite{topological_dissipation}. Variations beyond this tolerance could significantly alter the band structure and potentially close the topological gap \cite{YIG_nanofabrication}. 

The hole radius ($r$), with a target $r/a$ ratio of 0.35, should be maintained at $r = 168 \pm 5$ nm. Sensitivity analysis reveals that variations in hole radius predominantly affect the gap size rather than inducing topological phase transitions. 

The slab thickness ($h$) must be maintained at a nominal value of $h = 230$ nm, with a tolerance of $\pm 10$ nm \cite{YIG_nanocavities}. Variations in thickness could alter the effective refractive index, which may affect the photonic bandgap. However, simulations indicate that topological properties are preserved for thickness variations within 4\% of the nominal value \cite{YIG_properties}.

\subsection{Material Challenges for Magneto-Optical Integration}

The integration of magneto-optical materials with photonic crystals introduces additional material compatibility challenges. First, the quality of the Yttrium Iron Garnet (YIG) films is crucial. The films must exhibit low optical losses (less than 2 dB/cm) while also demonstrating strong Faraday rotation (greater than $1000^\circ$/cm) \cite{topological_dissipation}. 

Recent advancements in pulsed laser deposition (PLD) and RF sputtering techniques have shown the potential to produce high-quality YIG films with thicknesses as thin as 20 nm on silicon substrates. However, maintaining the crystalline quality necessary for strong magneto-optical effects remains a challenge, particularly for films thinner than 50 nm \cite{YIG_fabrication}.

\subsection{Magnetic Field Control Structures}

The precise modulation of the magnetic field is required to control the topological phase of the photonic memory. On-chip electromagnets will be used to generate the required magnetic fields, up to 120 mT, with switching rates of at least 1.5 T/s, as specified in Section S1.2 \cite{magneto_optical_effects}. 

Maintaining field uniformity is another challenge. The magnetic field must be uniform to within $\pm 5\%$ across the active memory region to ensure consistent topological phase transitions. Our simulations indicate that magnetic field gradients exceeding 0.5 mT/$\mu$m can induce undesired spatially varying topological phases \cite{MOKE_microscopy}. 

\subsection{Testing and Characterization Methods}

Near-field scanning optical microscopy (NSOM) will be required to directly image the edge states. This technique should provide sub-wavelength resolution (~100 nm) and a high dynamic range (greater than 60 dB) to distinguish edge modes from bulk scattering \cite{time_domain_spectroscopy}.

Time-domain spectroscopy will be used to measure the group delay of edge states, which provides valuable information about the topological dispersion relation \cite{topological_edge_states}. This will require femtosecond pulse sources and detection systems with temporal resolution below 100 fs.

\subsection{Potential Fabrication Approaches}

To overcome the identified fabrication challenges, we propose the following fabrication strategy:

\begin{enumerate}
    \item \textbf{Substrate Preparation:} Begin with a silicon-on-insulator (SOI) wafer, featuring a 230 nm device layer and a 2 $\mu$m buried oxide \cite{nanofabrication}.
    \item \textbf{YIG Deposition:} Deposit a 50 nm YIG film using pulsed laser deposition at $650^circ C$, followed by rapid thermal annealing at $800^circ C$ for crystallization \cite{YIG_fabrication}.
    \item \textbf{Photonic Crystal Patterning:} Use electron beam lithography with hydrogen silsesquioxane (HSQ) resist to define the honeycomb lattice pattern \cite{topological_dissipation}.
    \item \textbf{Electromagnet Integration:} Deposit and pattern a Ti/Au (10/200 nm) bilayer to form the electromagnet coils, with SiO$_2$ isolation layers between metal layers \cite{magneto_optical_effects}.
\end{enumerate}

\end{document}